\documentclass[twocolumn,superscriptaddress]{revtex4-1}
\usepackage{amsmath, bm, physics,amssymb}
\usepackage{color}
\usepackage{graphicx}
\usepackage{siunitx}
\usepackage{txfonts}

\usepackage[whole]{bxcjkjatype} 
\usepackage[colorlinks=true,urlcolor=blue,citecolor=blue,linkcolor=blue,breaklinks=true]{hyperref}
\input glyphtounicode
\newcommand{\smrm}[1]{_{\mathrm{#1}}}
\newcommand{\uprm}[1]{^{\mathrm{#1}}}

\usepackage{tikz}
\usetikzlibrary{positioning,shapes.misc}
\usepackage{tikz-feynhand}

\tikzset{
  dot/.style={
    circle,
    fill=black,
    minimum size=3pt,
    inner sep=0pt
  }
}
\begin{document}
\title{Nonreciprocal current induced by dissipation in time-reversal symmetric systems: relation to wave-packet skewness
}
\author{Takahiro Anan}
\affiliation{Department of Applied Physics, The University of Tokyo, Hongo, Tokyo,
  113-8656, Japan}
\affiliation{Department of Physics, Kyoto University, Kyoto,
  606-8502, Japan}  
\author{Sota Kitamura}
\affiliation{Department of Physics, Kyoto University, Kyoto,
  606-8502, Japan}
\author{Takahiro Morimoto}
\affiliation{Department of Physics, Kyoto University, Kyoto,
  606-8502, Japan}
\date{\today}
\begin{abstract}
We study nonreciprocal current response in noncentrosymmetric crystals under time-reversal symmetry.
Using the Green's function method, we derive the second-order DC conductivity at finite dissipation and express it in terms of the shift vector and a three-band geometric contribution.
The nonreciprocal current of the present mechanism turns out to be $O(1/\tau)$ ($\tau$: the lifetime of Bloch electrons).
Furthermore, in band-edge limits, the response reduces to a form governed by wave-packet skewness, encoded in the quantum geometric connection.
We present a numerical simulation of the nonreciprocal current in the one-dimensional Rice--Mele model and give its order estimation for nonmagnetic polar semiconductors.
\end{abstract}
\maketitle

\section{Introduction}
Nonreciprocity refers to the directional asymmetry of a physical response: the response to a drive in one direction is not equivalent to that for the opposite direction. 
This directional asymmetry originates from the breaking of inversion symmetry. 
At the same time, in the context of current response in crystals, microscopic time-reversal (TR) symmetry plays an essential role~\cite{Tokura2018,Nagaosa2024}.
This is because, within conventional Boltzmann transport theory, the longitudinal nonreciprocal current is described by the nonlinear Drude contribution, which requires a $k$-asymmetric band structure $\epsilon_k\neq\epsilon_{-k}$.
A well-known example of such nonreciprocal current response is the magnetochiral anisotropy~\cite{Rikken2001}, which requires a magnetic field or magnetization to realize the $k$-asymmetric band structure.
Accordingly, nonreciprocal current response in TR broken systems has been extensively studied both theoretically~\cite{Morimoto2016,Wakatsuki2018,Hoshino2018,Watanabe2020,Watanabe2021,Das2023,Kaplan2024} and experimentally~\cite{Ideue2017,Yokouchi2017,Wakatsuki2017,Yasuda2019,Itahashi2020,Zhang2020,Zhao2020,Nakamura2025,Li2021,Wang2022,Wakamura2024}.
While these studies are based on Bloch electrons with finite lifetime, there are several approaches to derive nonreciprocal current response in TR symmetric systems by incorporating additional effects.
For example, electron interactions can lead to a band modification dependent on the applied electric field, which induces nonreciprocal current \cite{Morimoto2018}.
Another example is skew scattering and side jump induced by inversion symmetry breaking, which provide $k$-asymmetric scattering processes and generate nonreciprocal current~\cite{Isobe2020,Varshney2026}.
This raises a basic question: can nonreciprocal current arise under TR symmetry in a noninteracting multiband system with a finite lifetime, without invoking these additional mechanisms?

Interband processes provide a physical route to nonreciprocal transport that is not captured by the band dispersion alone.
In the optical regime, the displacement of an electron wave packet during an interband excitation generates the shift current, whose geometric content is characterized by the shift vector~\cite{Sipe2000}.
A related displacement also enters nonreciprocal Landau--Zener tunneling under a strong DC electric field~\cite{Kitamura2020,Kitamura2020PRB}.
These effects can occur without breaking TR symmetry.
Optical excitation and strong-field tunneling, however, must be distinguished from the perturbative response to a weak static field.
The latter calls for a separate examination of how interband geometry enters the DC conductivity at finite dissipation.

The relaxation-time-independent contribution, of order $O(\tau^0)$ with $\tau$ the electronic lifetime, has been the subject of debate in longitudinal nonlinear DC transport.
Different formulations have yielded conflicting conclusions about the existence and geometric origin of a longitudinal current at this order~\cite{Michishita2022,Kaplan2023,Das2023,Wang2024,Kaplan2024,Xiao2025,Qiang2026,Tang2026,Ulrich2026,Guo2026}.
Recently, a study combining adiabatic perturbation theory with a nonequilibrium Green's function approach has clarified the differences among these formulations and demonstrated the presence of a longitudinal quantum metric dipole contribution at this order~\cite{Kitamura2026Quasiequilibrium}.
However, the quantum metric dipole contribution requires TR symmetry breaking, as well as the nonlinear Drude contribution.
Thus, the nonreciprocal current and its geometric origin in TR-symmetric systems have not been fully clarified.

\begin{figure}
  \includegraphics[width=\linewidth]{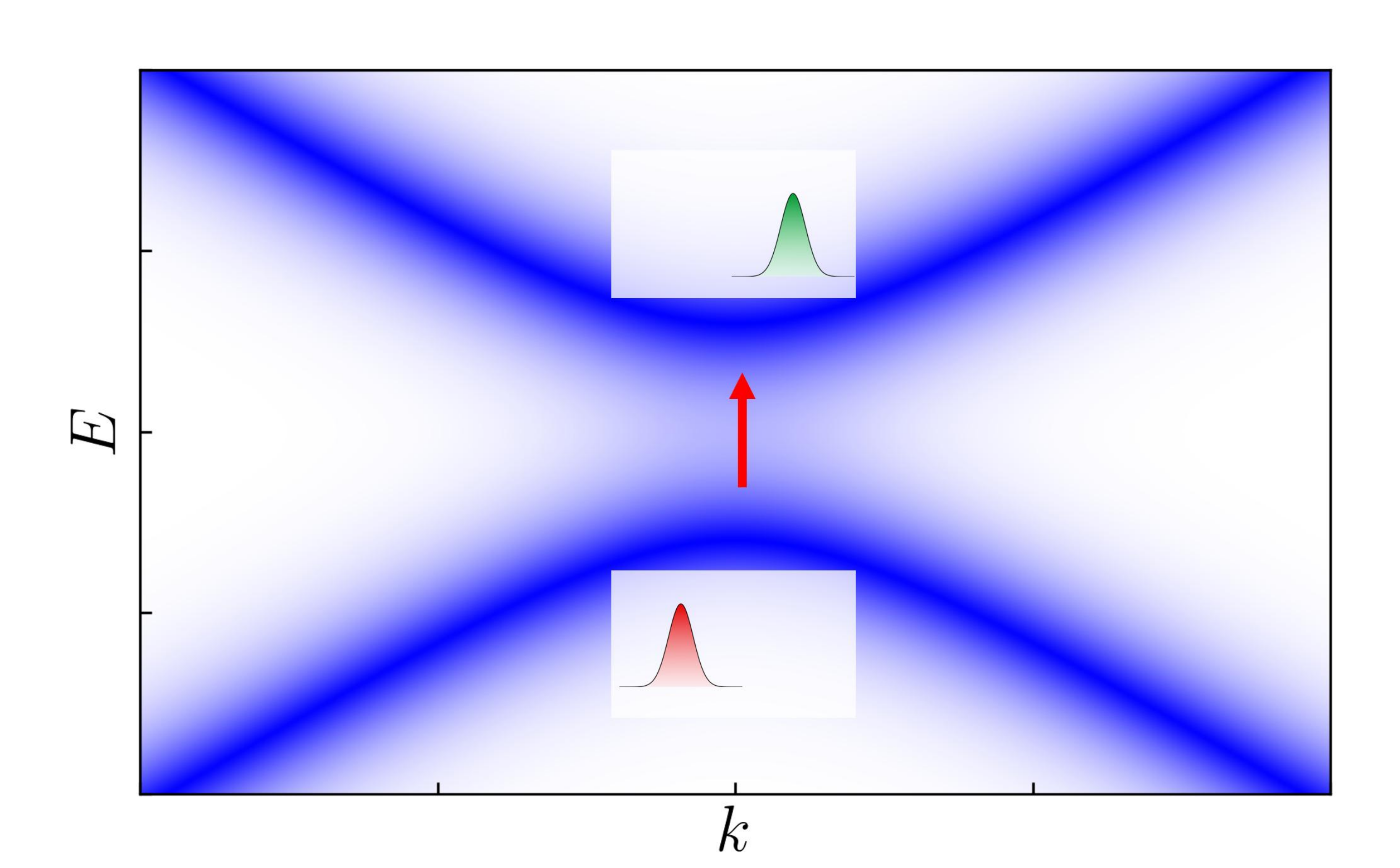}
  \caption{Nonreciprocal current induced by dissipation. 
  With finite dissipation, Bloch electrons can be excited to upper bands upon application of an electric field, which produces nonreciprocal current in inversion symmetry broken systems.
  }
  \label{fig:schematic}
\end{figure}

Here, we show that nonreciprocal current emerges even in TR symmetric systems when interband transitions enabled by a finite relaxation rate are taken into account.
We derive a formula for the nonreciprocal current in TR symmetric systems in the presence of dissipation in a gauge-invariant form and show that it arises from the electron wave packet shift in the interband transition exemplified by the shift vector $R$ (Fig.~\ref{fig:schematic}).
Particularly, in the band-edge limit, the nonreciprocal current is expressed in terms of the wave-packet skewness, which corresponds to the imaginary part of the quantum geometric connection.
We also perform numerical calculations using the Rice--Mele model to investigate the behavior of the nonreciprocal current.

\section{Results}
We derive the nonreciprocal current as a static limit of the second order conductivity $\mathcal{K}^{\mu\alpha\beta}(\omega_1,\omega_2)$ that is defined as 
\begin{align}
j_\mu^{(2)}(t)=&\sum_{\alpha,\beta}\int\frac{d\omega_1}{2\pi}\int \frac{d\omega_2}{2\pi}\nonumber \\
  &\quad \times \mathcal{K}^{\mu\alpha\beta}(\omega_1,\omega_2)A_\alpha(\omega_1)A_\beta(\omega_2)e^{-i(\omega_1+\omega_2)t}, \label{eq:definitionOfCurrent}
\end{align}
where $j_\mu^{(2)}(t)$ is the nonlinear current density and $\mu,\alpha,\beta$ are the spatial indices.
We consider a monochromatic electric field with the velocity gauge $E(t)=-\partial_t A(t),\ A_\alpha(\omega_1)=A_\alpha 2\pi\delta(\omega_1-\omega) +A_\alpha^* 2\pi\delta(\omega_1+\omega)$.
In this formalism, the static nonlinear conductivity $j_\mu^{(2)}(t)=\sigma^{\mu\alpha\alpha} E_\alpha^2(t)$ is given as an $O(\omega^2)$ term of $\mathcal{K}^{\mu\alpha\alpha}$, namely,
\begin{align}
  \sigma^{\mu\alpha\alpha}=& \frac{1}{2}\partial_\omega^2\mathcal{K}^{\mu\alpha\alpha}(\omega,-\omega)|_{\omega=0}. \label{eq:nonReciprocalEsquared}
\end{align}
We refer to the longitudinal component $\sigma^{\alpha\alpha\alpha}$ as the nonreciprocal conductivity, since, for a static electric field along the $\alpha$ direction, the nonreciprocal current is given as $(j_\alpha(E_\alpha)+j_\alpha(- E_\alpha))/2=\sigma^{\alpha\alpha\alpha} E_\alpha^2+O(E_\alpha^4)$.
Experimentally, this nonreciprocity is measured as the second-harmonic resistance or voltage, as shown in Appendix~\ref{app:orderEstimation}.
Here, $O(\omega^0)$ components and $O(\omega)$ components of $\mathcal{K}^{\mu\alpha\alpha}(\omega,-\omega)$ vanish because of the gauge invariance.
The detailed derivations are shown in Appendix~\ref{app:deriveDC}.

Using the Green function method, $\mathcal{K}^{\mu\alpha\beta}(\omega,-\omega)$ is given by
\begin{align}
  \mathcal{K}^{\mu\alpha\beta}(\omega,-\omega)=&\left(\frac{e}{\hbar}\right)^3\int \frac{d\epsilon}{2\pi i}\int_{\mathrm{BZ}} \frac{d^d k}{(2\pi)^d} 2\Gamma^2 (f(\epsilon)-f(\epsilon+\hbar\omega)) \nonumber \\
  \times \mathrm{Tr} [\partial_{k_\mu} (G^R&(\epsilon+\hbar\omega) H^\alpha G^A(\epsilon) )G^R(\epsilon) H^\beta G^A(\epsilon+\hbar\omega)] ,\label{eq:currentbyExcitation}
\end{align}
where $f(\epsilon)=(e^{\beta \epsilon}+1)^{-1}$ is the Fermi distribution function, $H^\alpha=\partial_{k_\alpha} H$ with $H$ being the Hamiltonian, and $\mathrm{Tr}$ denotes the trace over band indices.
Here, $G^{R(A)}(\epsilon)=(\epsilon+\mu-H\pm i\Gamma)^{-1}$ is the retarded (advanced) Green's function with the relaxation rate $\Gamma$, and we introduce this relaxation rate phenomenologically to describe the dissipation in the system.
This assumption is justified when the fermionic bath in equilibrium has a constant density of states.
Under this assumption, the vertex correction vanishes, and thus the photovoltaic effect with the finite relaxation rate $\Gamma$ is given by Eq.~\eqref{eq:currentbyExcitation} without vertex corrections.
In the clean limit $\Gamma \to 0$, Eq.~\eqref{eq:currentbyExcitation} reproduces the expressions for nonlinear optical conductivities such as injection and shift current \cite{Sipe2000}. 
In contrast, taking the static limit $\omega \to 0$ before the clean limit $\Gamma \to 0$ yields the expressions for the nonreciprocal current responses including the nonlinear Drude term, the Berry curvature dipole term, and the quantum metric dipole term \cite{Ulrich2026}.
The derivations of these limiting cases are given in Appendices~\ref{app:injectionShift} and~\ref{app:DrudeBCD}, respectively.

Here, we consider the nonreciprocal current (i.e. $\mu=\alpha$ in Eq.~\eqref{eq:nonReciprocalEsquared}) in TR symmetric systems.
The nonreciprocal current is given by
\begin{align}
  \sigma^{\alpha\alpha\alpha}=& \frac{e^3}{\hbar} \int\smrm{BZ}\frac{d^d k}{(2\pi)^d}\left[\sum_{a\neq b}R_{ab}|A_{ab}|^2D^{(2)}_{ab}\right.\nonumber \\
  &\left.+\sum_{a\neq b,b\neq c,c\neq a}\Re[A_{ab}A_{bc}A_{ca}]D^{(3)}_{abc} \right] \label{eq:sigma2}
\end{align}
with
\begin{align}
  D^{(2)}_{ab}&=\Re \left[\frac{8\Gamma (\Delta_{ab}+i\Gamma)}{(\Delta_{ab}+2i\Gamma)^2}f_{+,a}'-\frac{2\Gamma \Delta_{ab}}{\Delta_{ab}+2i\Gamma}f_{+,a}''\right] \\
  D^{(3)}_{abc}&=\Re\left[-\left(\frac{1}{\Delta_{ac}}+\frac{1}{\Delta_{bc}}\right)\frac{8\Gamma\Delta_{ab}}{\Delta_{ab}+2i\Gamma}f_{+,a}'+\frac{2\Gamma\Delta_{ab}}{\Delta_{ab}+2i\Gamma}f_{+,a}''\right]
\end{align}
where $A_{ab}=i\bra{u_a}\partial_{k_\alpha} \ket{u_b}$ is the interband Berry connection, $\Delta_{ab}=\epsilon_a-\epsilon_b$ is the band gap, $R_{ab}=\Im \partial_{k_\alpha}( \log A_{ba})+i\bra{u_a}\partial_{k_\alpha}\ket{u_a}-i\bra{u_b}\partial_{k_\alpha}\ket{u_b}$ is the shift vector, $f_{\pm}(x)\equiv \frac{1}{4}\pm \frac{1}{2\pi i}\psi\left(\frac{1}{2}\pm \frac{\beta x}{2\pi i}\right)$ with $\psi(z)$ being the digamma function, and $f_{+,a}'=f_+'(\epsilon_a-\mu+i\Gamma),\ f_{+,a}''=f_+''(\epsilon_a-\mu+i\Gamma)$.
Here, we assume that there is no degeneracy in the band structure for simplicity.
Under independent phase transformations of the Bloch eigenstates, $\ket{u_a}\to e^{i\phi_a(\bm{k})}\ket{u_a}$, the two geometric factors $R_{ab}|A_{ab}|^2$ and $\Re[A_{ab}A_{bc}A_{ca}]$ are each gauge invariant.
Consequently, Eq.~\eqref{eq:sigma2} is manifestly gauge invariant.
The detailed derivation of Eq.~\eqref{eq:sigma2} is shown in Appendix~\ref{app:derivationSigma2}.
The coefficients $D^{(2)}_{ab}$ and $D^{(3)}_{abc}$ vanish at least linearly with $\Gamma$, indicating that the nonreciprocal conductivity vanishes in the clean limit.
The leading-order dependence on $\Gamma$, which varies with temperature and the position of the chemical potential, is discussed in Appendix~\ref{app:leadingOrderGamma}.

We now examine limiting regimes that reveal the geometric nature of Eq.~\eqref{eq:sigma2}.
We assume that the bands are well separated $|\Delta_{ab}|>\Delta\smrm{min}\gg \Gamma$ for all $a\neq b$.
In this case, we can expand $D^{(2)}_{ab}$ and $D^{(3)}_{abc}$ in terms of $\Gamma$ as $D^{(2)}_{ab}=\Gamma \left(\frac{4}{\Delta_{ab}}f_a'-f_a''\right)+O(\Gamma^2),\ D^{(3)}_{abc}=\Gamma \left[-4\left(\frac{1}{\Delta_{ac}}+\frac{1}{\Delta_{bc}}\right)f_a'+f_a''\right]+O(\Gamma^2)$
Next, we consider the low-temperature regime in which the chemical potential is offset from the band edge by an energy of order $\beta^{-1}$, i.e. $\Delta\smrm{min}\gg |\xi_a|\sim \beta^{-1}(\gg \Gamma)$.
In this situation, we can ignore the first term in $D^{(2)}_{ab}$ and $D^{(3)}_{abc}$, and we obtain
\begin{align}
  D^{(2)}_{ab}\simeq -\Gamma f''_{a},\
  D^{(3)}_{abc}\simeq \Gamma f''_{a}.
\end{align}
By substituting these expressions into Eq.~\eqref{eq:sigma2}, we obtain
\begin{align}
  \sigma^{\alpha\alpha\alpha}\simeq \frac{e^3}{\hbar}\int_{\mathrm{BZ}}\frac{d^d k}{(2\pi)^d}\sum_{a}^{|\xi_a|\sim \beta^{-1}}\Gamma f''_a \Re \mel{u_a}{(i\partial_{k_\alpha}-A_{aa})^3}{u_a} \label{eq:skewness}
\end{align}
Here, the gauge-invariant quantity $\Re \mel{u_a}{(i\partial_{k_\alpha}-A_{aa})^3}{u_a}$ characterizes the contribution of Bloch-band geometry to the real-space skewness of an electron wave packet,
\begin{align}
  \Re \mel{u_a}{(i\partial_{k_\alpha}-A_{aa})^3}{u_a}\sim \mel{u_a}{(\Delta r_\alpha)^3}{u_a}.
\end{align}
This quantity is odd in momentum in the presence of inversion symmetry and even in momentum in the presence of TR symmetry. 
It therefore vanishes identically when both symmetries are present. 
Consequently, a nonzero value provides a geometric signature of inversion-symmetry breaking in the TR-symmetric systems considered here.
This quantity is  also the imaginary part of the quantum geometric connection, 
\begin{align}
  \Re \mel{u_a}{(i\partial_{k_\alpha}-A_{aa})^3}{u_a}=-\Im C_a^{\alpha\alpha\alpha}
\end{align} 
where $C_a^{\alpha\beta\gamma}\equiv \Tr[P_a \partial_\beta P_a \partial_\alpha \partial_\gamma P_a]$~\cite{Mitscherling2025}.
Therefore, Eq.~\eqref{eq:skewness} indicates that the nonreciprocal current is induced by this skewness, which is a geometric quantity.
Although this is valid for the regime $\Gamma \ll \beta^{-1} \ll \Delta\smrm{min}$, we can obtain similar expressions for the regime $\beta^{-1} \ll \Gamma \ll \Delta\smrm{min}$, which is given as 
\begin{align}
  \sigma^{\alpha\alpha\alpha}\simeq \frac{e^3}{\hbar}\int_{\mathrm{BZ}}\frac{d^d k}{(2\pi)^d}\sum_{a}^{|\xi_a|\sim \Gamma}\Gamma \frac{d n_a}{d \mu} \Re \mel{u_a}{(i\partial_{k_\alpha}-A_{aa})^3}{u_a}, \label{eq:skewness-low-temp}
\end{align}
where $n_a=\Gamma / [\pi (\xi_a^2+\Gamma^2)]$ is the spectral weight of the $a$-th band.
The detailed derivation of these limiting cases is shown in Appendix~\ref{app:skewnessLowTemp}.
Note that Eqs.~\eqref{eq:skewness} and \eqref{eq:skewness-low-temp} has a finite value when the chemical potential lies near the band edge after integrating over the Brillouin zone.

\begin{figure}
  \includegraphics[width=\linewidth]{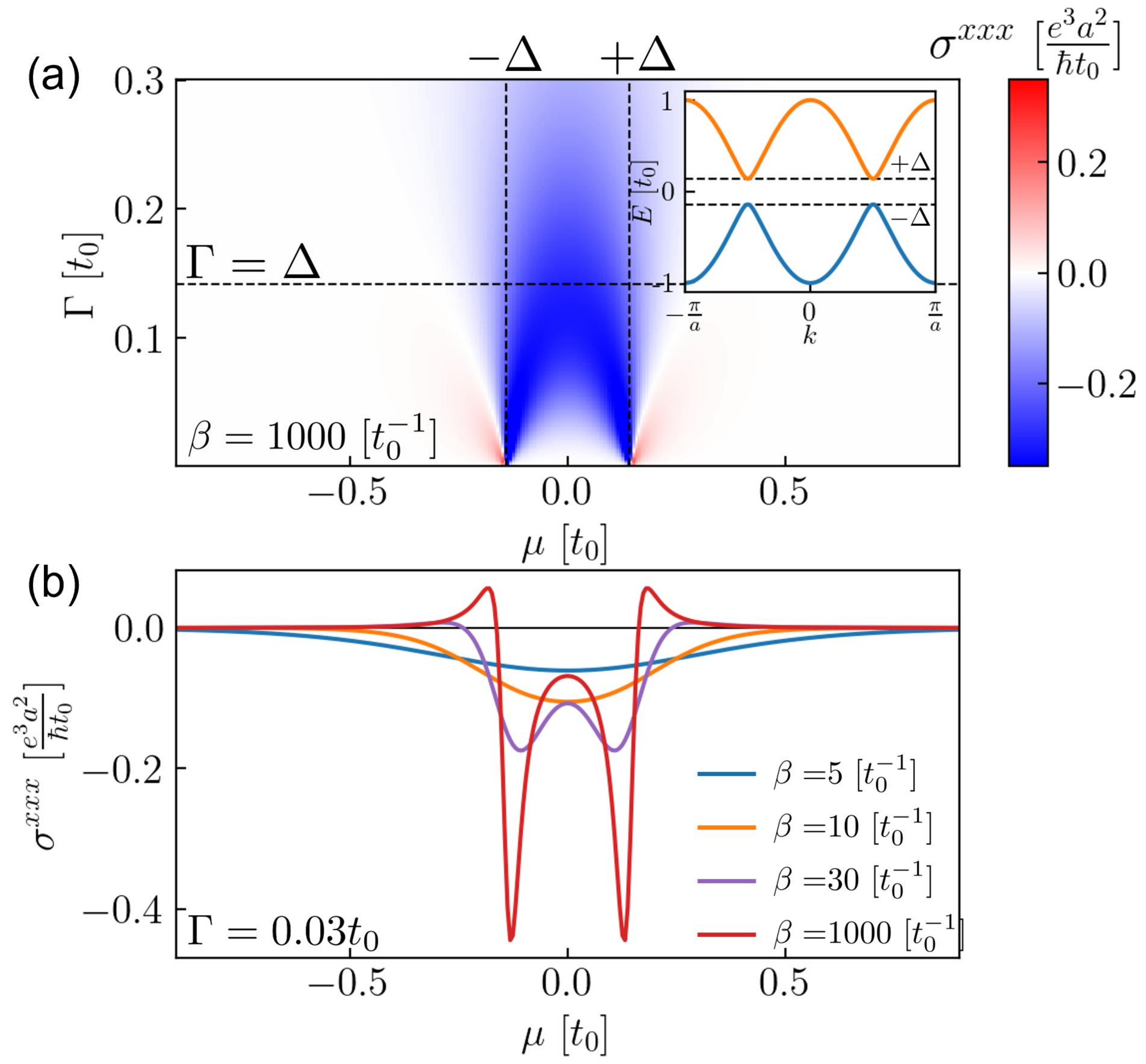}
  \caption{Nonreciprocal current $\sigma^{xxx}$ of the Rice--Mele model. (a) Color plot of $\sigma^{xxx}$ as a function of chemical potential $\mu$ and relaxation rate $\Gamma$. Dashed lines indicate the half gap $\Delta=\sqrt{m^2+\delta t^2}$. The inset in (a) shows the band structure of the Rice--Mele model. (b) $\sigma^{xxx}$ as a function of $\mu$ for different values of $\beta$. We set the parameters as $m=0.1t_0,\delta t=0.1t_0$.}
  \label{fig:colorMapvsMu}
\end{figure}

\section{Model calculation}
To demonstrate the nonreciprocal current induced by dissipation, we consider the Rice--Mele model given by $\mathcal{H}(k)=t_0\cos ka \sigma_x + \delta t \sin ka \sigma_y + m\sigma_z$, where $a$ is the lattice constant and $\sigma_{x,y,z}$ are the Pauli matrices.
The Rice--Mele model is a minimal model for ferroelectric materials and contains two inversion-symmetry-breaking parameters: the staggered onsite potential $m$ and the staggered hopping amplitude $\delta t$. 
Because the Rice--Mele model preserves TR symmetry, the nonreciprocal current is not induced by the nonlinear Drude term or the quantum metric dipole term.
Therefore, in this model, the nonreciprocal current arises only from the dissipation-induced nonreciprocal current described by Eq.~\eqref{eq:sigma2}.

Figure~\ref{fig:colorMapvsMu} (a) shows the nonreciprocal current $\sigma^{xxx}$ of the Rice--Mele model as a function of chemical potential $\mu$ and relaxation rate $\Gamma$.
The inset in Fig.~\ref{fig:colorMapvsMu} (a) shows the band structure of the Rice--Mele model which has band edges at $E = \pm \Delta$ with $\Delta/t_0 \simeq 0.14$.
The nonreciprocal current is enhanced when the chemical potential is near the band gap $-\Delta \lesssim \mu \lesssim \Delta$ and the relaxation rate is comparable to the band gap $\Gamma \sim \Delta$.
This is consistent with the fact that the nonreciprocal current is induced by the interband excitation.
Figure~\ref{fig:colorMapvsMu} (b) shows $\sigma^{xxx}$ as a function of $\mu$ for different values of $\beta$.
In the low temperature regime $\beta \Gamma \gg 2\pi$, the nonreciprocal current shows peaks and sign changes near the band edge.
This behavior can be understood from Eqs.~\eqref{eq:skewness} and \eqref{eq:skewness-low-temp}, in which the skewness of the wave function is weighted by $\Gamma f_a''$ for $\Gamma \ll \beta^{-1} \ll \Delta\smrm{min}$ and by $\Gamma\, d n_a/d\mu$ for $\beta^{-1} \ll \Gamma \ll \Delta\smrm{min}$ inside the Brillouin-zone integral, which does not cancel out when the chemical potential is near the band edge.
In the high temperature regime $\beta \Gamma \ll 2\pi$, the nonreciprocal current shows a single peak in the band gap, as the difference of the Fermi distribution function of the upper and lower bands is enhanced in the band gap. 
Similar behavior is observed in the nonreciprocal current of the three-band model, as shown in Appendix~\ref{app:threeBand}.

\begin{table*}[t]
  \begin{tabular}{c|c|c|c}
  \hline\hline
     $\sigma^{\alpha\alpha\alpha}$&TR symmetric systems &Relaxation time ($\tau$) dependence & Inter/intra-band mechanism\\
    \hline
    Nonlinear Drude term & - &$O(\tau^{2})$ &Intraband \\
    Quantum metric dipole term& - &  $O(\tau^{0})$ &Intraband\\
    Skew scattering and side jump & $\checkmark$ & $O(\tau^{2}) \text{or} O(\tau)$ &Intraband \\
    Dissipation induced nonreciprocity (present) & $\checkmark$ & $O(\tau^{-1})$ &Interband \\
    \hline\hline
  \end{tabular}
  \caption{
  Comparison of nonreciprocal currents in noninteracting systems.
  The nonlinear Drude and quantum metric dipole terms require time-reversal (TR) symmetry breaking, whereas both the disorder-induced contributions from skew scattering and side jump and the present contribution are allowed in TR-symmetric systems. 
  The present contribution is distinguished from the skew-scattering and side-jump contributions by its interband character and its vanishing behavior in the clean limit $\tau \to \infty$.
  The expressions for the nonlinear Drude and quantum metric dipole terms are given in Secs.~\ref{app:NLD} and \ref{app:QMD} of the Appendix.
  }
  \label{tab:comparison}
\end{table*}

\section{Discussion}
We have demonstrated that nonreciprocal current can be induced by dissipation in TR symmetric systems.
In this section, we discuss the characteristics of this nonreciprocal current, an order estimate and candidate materials for its observation.
Table~\ref{tab:comparison} compares the present dissipation-induced contribution with previously proposed mechanisms for nonreciprocal current in noninteracting systems.
The present contribution has a leading term of order $O(\tau^{-1})$, whereas the other contributions listed in the table scale as $O(\tau^2)$, $O(\tau)$, or $O(\tau^0)$.
It therefore vanishes in the clean limit and is enhanced by dissipation in the weak-relaxation regime.
Moreover, the present contribution is allowed in TR-symmetric systems, where the nonlinear Drude and quantum metric dipole terms vanish.
Its interband origin further distinguishes it from the TR-symmetric intraband contributions of skew scattering and side jump.
The contrasting relaxation-time dependences provide a possible experimental means of distinguishing these contributions in the weak-relaxation regime.
A complementary test is to tune the chemical potential near a band edge and examine the peaks and sign changes predicted at low temperatures in Fig.~\ref{fig:colorMapvsMu}(b), which would capture the quantum geometric connection or skewness of the wave function in Eqs.~\eqref{eq:skewness} and \eqref{eq:skewness-low-temp}.
Comparing these transport measurements with independent optical shift-current measurements across the relevant band gap may further help assess the role of the shift-vector contribution in the nonreciprocal current.

Another characteristic of the nonreciprocal current induced by dissipation is that it can be observed even when the chemical potential $\mu$ lies in the band gap, in contrast to the previous mechanisms for nonreciprocal current in TR symmetric systems, which require the chemical potential to lie within a band~\cite{Morimoto2018,Isobe2020,Varshney2026}.
From an experimental perspective, nonreciprocal signals such as the second-harmonic voltage $V^{2\omega}\propto I^2\sigma^{xxx}/(\sigma^{xx})^3$ and the second-harmonic resistance $R^{2\omega} \propto I\sigma^{xxx}/(\sigma^{xx})^3$ are often measured as functions of the applied current $I$. 
Therefore, the present mechanism is suitable for materials with low conductivity, in particular wide-gap semiconductors with low carrier density and long relaxation times. 
We note that this situation does not necessarily mean large $\sigma^{xxx}$ itself which is enhanced for short relaxation times ($\Gamma \sim \Delta$).

In the present analysis, we consider that all quasiparticles have the same relaxation time $\tau=\hbar/(2\Gamma)$, since the relaxation is incorporated in the constant self energy in the Green's function. 
This assumption is well justified when the system is coupled to a fermionic bath with a constant density of states.
However, even in the situation where the relaxation rate $\Gamma$ originates from impurity scattering or related mechanisms, our analysis is applicable to a situation where the quasiparticle states are not localized, but lie on the extended-state side of the mobility edge.
In these situations, the vertex corrections may change the quantitative value of the nonreciprocal current, but the qualitative behavior of the nonreciprocal current is expected to be unchanged.
With the requirement of extended quasiparticle states, three-dimensional doped semiconductors are promising candidate materials for observing the present nonreciprocal current.
We note that the localization effect is not incorporated in the present analysis and is out of scope for this study.

To estimate these nonreciprocal signals in physical units, we derive $\sigma^{xx}$ with finite relaxation rate $\Gamma$ and perform an order estimate of $V^{2\omega}$ and $R^{2\omega}$ for a three-dimensional extension of the Rice--Mele model in Appendix~\ref{app:orderEstimation}.
For a sample with cross-sectional area $S=1\,\mathrm{mm}^2$ and length $l=1\,\mathrm{mm}$, a representative semiconductor with a band gap $2\Delta=3.1\,\mathrm{eV}$ and a relaxation time of $0.3\,\mathrm{ps}$ gives $|R^{2\omega}|\sim 0.3\,\Omega$ and $|V^{2\omega}|\sim 0.02\,\mathrm{mV}$ for a dissipated power of $1\,\mathrm{mW}$ and an applied current of $0.05\,\mathrm{mA}$.
These nonreciprocal signals are comparable to or larger than the experimentally observed nonreciprocal signals in TR broken systems~\cite{Ideue2017,Yokouchi2017,Wakatsuki2017,Yasuda2019,Itahashi2020,Zhang2020,Nakamura2025,Li2021,Wang2022,Wakamura2024}.
Moreover, these nonreciprocal signals are observable at room temperature, which is an advantage for potential applications.
The detailed estimation is shown in Appendix~\ref{app:orderEstimation}.

Based on the above discussion, promising candidate materials are nonmagnetic polar semiconductors with sufficiently large band gaps, including visible- and ultraviolet-gap materials, and relatively long relaxation times.
In addition, a second-order longitudinal response $\sigma^{\alpha\alpha\alpha}$ should be allowed by crystal symmetry.
Since the present nonreciprocal current contains a contribution governed by the shift vector $R$, materials in which shift current has already been experimentally observed are particularly attractive candidates for the observation of the present nonreciprocal current.
Wurtzite semiconductors such as CdS and CdSe are representative examples~\cite{Laman2005,Sotome2021}.
Ferroelectric semiconductors such as SbSI are similarly attractive because their band gaps lie near the visible range and their shift-current response can be switched by reversing the ferroelectric polarization~\cite{Ogawa2017,Sotome2019}.
Ferroelectric oxides such as BaTiO$_3$~\cite{Koch1975,Okamura2022}, LiNbO$_3$~\cite{Glass1974,Dalba1995} and Pb-based ferroelectric oxides~\cite{Yang2000,Poosanaas1998,Pintilie2007} are other candidate materials.
Even in nonmagnetic polar semiconductors where a shift-current response has not yet been experimentally established, nonmagnetic polar semiconductors with sufficiently large band gaps can serve as candidate platforms.
For example, ZnO, GaN, AlN, and AlGaN alloys in the ideal wurtzite phase are noncentrosymmetric polar semiconductors with point group $C_{6v}$, and therefore allow a longitudinal response $J_c\propto E_c^2$ along the polar $c$ axis~\cite{Vurgaftman2003,Ozgur2005}.

\begin{acknowledgements}
We thank Naoto Nagaosa, Yannis Ulrich, and Yoshihiro Iwasa for insightful discussions. This work was supported by JST SPRING, Grant Number JPMJSP2108 (TA) and MEXT/JSPS KAKENHI, Grant Numbers JP26KJ0822 (TA), JP25H01249, JP25K07219 (SK), JP24H02231, JP23K17665, JP24K00568 (TM), JP23K25816 (TM, SK).
\end{acknowledgements}

\appendix

\section{Derivation of the DC response} \label{app:deriveDC}
In this section, we show that the nonreciprocal current is given by Eq.~\eqref{eq:nonReciprocalEsquared} and show that there is no divergence in the static limit of the second order conductivity $\sigma^{\mu\alpha\beta}(\omega_1,\omega_2)$ with respect to the electric field.

\subsection{Derivation of the DC response from the AC response}
First, we show that the nonreciprocal current is given by Eq.~\eqref{eq:nonReciprocalEsquared}.
Using $A_\alpha(\omega_1)=A_\alpha 2\pi\delta(\omega_1-\omega) +A_\alpha^* 2\pi\delta(\omega_1+\omega)$, we can rewrite the current density (Eq.~\eqref{eq:definitionOfCurrent}) as
\begin{align}
  j_\mu(\omega)\equiv &\int_{-\infty}^{\infty}dt j_\mu(t) e^{i\omega' t}\nonumber\\
  =& j_\mu\uprm{DC}(\omega)2\pi \delta(\omega')+j_\mu\uprm{SHG+}(\omega)2\pi\delta(\omega'-2\omega)\nonumber \\
  &+j_\mu\uprm{SHG-}(\omega)2\pi\delta(\omega'+2\omega)\nonumber \\
  j_\mu\uprm{DC}(\omega)=&\sum_{\alpha,\beta}\mathcal{K}^{\mu\alpha\beta}(\omega,-\omega) A_\alpha(\omega) A_\beta(-\omega)\nonumber\\
  &+\sum_{\alpha,\beta}\mathcal{K}^{\mu\alpha\beta}(-\omega,\omega) A_\alpha(-\omega) A_\beta(\omega) 
   \label{eq:DCsumOfalpha_beta}\\
  j_\mu\uprm{SHG+}(\omega)=&\sum_{\alpha,\beta}\mathcal{K}^{\mu\alpha\beta}(\omega,\omega) A_\alpha(\omega) A_\beta(\omega)\\
  j_\mu\uprm{SHG-}(\omega)=&\sum_{\alpha,\beta}\mathcal{K}^{\mu\alpha\beta}(-\omega,-\omega) A_\alpha(-\omega) A_\beta(-\omega).
\end{align}
Using $\bm{A}(\omega)=\bm{E}(\omega)/i\omega$, we can expand $j_\mu(t)$ in terms of $\omega$ as
\begin{widetext}
\begin{align}
  &j_\mu(t)\nonumber \\
  =&j_\mu\uprm{DC}(\omega)+j_\mu\uprm{SHG+}(\omega)e^{-i2\omega t}+j_\mu\uprm{SHG-}(\omega)e^{i2\omega t}\nonumber \\
  =&\sum_{\alpha,\beta}\frac{E_\alpha E_\beta^*}{\omega^2}
    (\mathcal{K}^{\mu\alpha\beta}|_{\omega_1=0,\omega_2=0}+\omega(\partial_{\omega_1}-\partial_{\omega_2})\mathcal{K}^{\mu\alpha\beta}|_{\omega_1=0,\omega_2=0}+\frac{1}{2}\omega^2 (\partial_{\omega_1}-\partial_{\omega_2})^2\mathcal{K}^{\mu\alpha\beta}|_{\omega_1=0,\omega_2=0}+\dots) \nonumber \\
  &+\sum_{\alpha,\beta}\frac{E_\alpha^* E_\beta}{\omega^2}(\mathcal{K}^{\mu\alpha\beta}|_{\omega_1=0,\omega_2=0}-\omega(\partial_{\omega_1}-\partial_{\omega_2})\mathcal{K}^{\mu\alpha\beta}|_{\omega_1=0,\omega_2=0}+\frac{1}{2}\omega^2 (\partial_{\omega_1}-\partial_{\omega_2})^2\mathcal{K}^{\mu\alpha\beta}|_{\omega_1=0,\omega_2=0}+\dots)\nonumber \\
  &-\sum_{\alpha,\beta}\frac{E_\alpha E_\beta}{\omega^2}
    (\mathcal{K}^{\mu\alpha\beta}|_{\omega_1=0,\omega_2=0}+\omega(\partial_{\omega_1}+\partial_{\omega_2})\mathcal{K}^{\mu\alpha\beta}|_{\omega_1=0,\omega_2=0}+\frac{1}{2}\omega^2 (\partial_{\omega_1}+\partial_{\omega_2})^2\mathcal{K}^{\mu\alpha\beta}|_{\omega_1=0,\omega_2=0}+\dots) e^{-i2\omega t}\nonumber \\
    & -\sum_{\alpha,\beta}\frac{E_\alpha^* E_\beta^*}{\omega^2} (\mathcal{K}^{\mu\alpha\beta}|_{\omega_1=0,\omega_2=0}-\omega(\partial_{\omega_1}+\partial_{\omega_2})\mathcal{K}^{\mu\alpha\beta}|_{\omega_1=0,\omega_2=0}+\frac{1}{2}\omega^2 (\partial_{\omega_1}+\partial_{\omega_2})^2\mathcal{K}^{\mu\alpha\beta}|_{\omega_1=0,\omega_2=0}+\dots)  e^{i2\omega t} \label{eq:expandK}
\end{align}
\end{widetext}
In the above expansion, only terms containing derivatives with respect to both $\omega_1$ and $\omega_2$ can survive.
Terms lacking such derivatives would yield a finite response to a static vector potential, in contradiction with gauge invariance.
In a Bloch-band description, a uniform static vector potential corresponds to the momentum shift $\bm{k}\to \bm{k}+\frac{e}{\hbar}\bm{A}$ applied consistently to the distribution function, band dispersion, and wave functions, and therefore does not change physical observables.
Therefore, within a theory based on Bloch's theorem, terms not differentiated with respect to $\omega_{1(2)}$ are expected to reduce to total derivative terms with respect to $\partial_{\alpha(\beta)}$.
Indeed, 
\begin{gather}
\mathcal{K}^{\mu\alpha\beta}(\omega_1,\omega_2)|_{\omega_1=0,\omega_2=0}=0,\\ \partial_{\omega_{1(2)}}\mathcal{K}^{\mu\alpha\beta}(\omega_1,\omega_2)|_{\omega_1=0,\omega_2=0}=0,\\ \partial_{\omega_{1(2)}}^2\mathcal{K}^{\mu\alpha\beta}(\omega_1,\omega_2)|_{\omega_1=0,\omega_2=0}=0,     
\end{gather}
as shown in the next subsection (Appendix~\ref{app:noDivergence}).

Therefore, the nonreciprocal current (i.e. the $O(\omega^0)$ term in Eq.~\eqref{eq:expandK}) is given as
\begin{align}
  j_\mu(t)=&\sum_{\alpha,\beta}(E_\alpha E_\beta^* +E_\alpha^* E_\beta +E_\alpha E_\beta e^{-i2\omega t}+E_\alpha^* E_\beta^* e^{i2\omega t})\nonumber \\
  & \qquad \times(-\partial_{\omega_1}\partial_{\omega_2}\mathcal{K}^{\mu\alpha\beta}|_{\omega_1=0,\omega_2=0} +O(\omega)) \nonumber \\
  =&\sum_{\alpha,\beta}(E_\alpha e^{-i\omega t}+c.c.) (E_\beta e^{-i\omega t}+c.c.)\nonumber \\
  & \qquad \times(-\partial_{\omega_1}\partial_{\omega_2}\mathcal{K}^{\mu\alpha\beta}|_{\omega_1=0,\omega_2=0} +O(\omega))  \nonumber \\
  =&\sum_{\alpha,\beta}E_\alpha(t)E_\beta(t)(-\partial_{\omega_1}\partial_{\omega_2}\mathcal{K}^{\mu\alpha\beta}|_{\omega_1=0,\omega_2=0} +O(\omega)) .
  \label{eq:11ee}
\end{align}
This is, for example, written by using photovoltaic conductivity $\mathcal{K}^{\mu\alpha\beta}(\omega,-\omega)$ as
\begin{align}
  j_\mu(t)=\sum_{\alpha,\beta}E_\alpha(t)E_\beta(t)\left(\frac{1}{2}\partial_{\omega}^2\mathcal{K}^{\mu\alpha\beta}(\omega,-\omega)|_{\omega=0} +O(\omega)\right)
\end{align}
When only $E_\alpha$ is nonzero among the components of the electric field $\bm{E}$, the nonreciprocal current is given by Eq.~\eqref{eq:nonReciprocalEsquared}.

\subsection{No divergence in the static limit} \label{app:noDivergence}
The response function $\mathcal{K}^{\mu\alpha\beta}(\omega_1,\omega_2)$ at the Matsubara frequency is given as
\begin{align}
&\mathcal{K}^{\mu\alpha\beta}(i\omega_1,i\omega_2)
\nonumber\\
=&-\frac{1}{2}\left(\frac{e}{\hbar}\right)^3
\begin{tikzpicture}[baseline=(b.center)]
  \begin{feynhand}
    \vertex (a) at (0,0);
    \vertex (b) at (1,0);
    \propag[fermion] (a) to[half left]    node[midway,above] {$i\omega',a$} (b);
    \propag[plain] (b) to[half left] (a);
    \node[dot] at (b) {};
    \propag[photon]  (b) to ++(1, 1)     node[left]  {$\alpha,i\omega_1$};
    \propag[photon]  (b) to ++(1,-1)     node[left]  {$\beta,i\omega_2$};
    \propag[photon]  (b) to ++( 1, 0)     node[right] {$\mu,i\omega_1+i\omega_2$};
  \end{feynhand}
\end{tikzpicture}
\;\nonumber\\ 
&-\frac{1}{2}\left(\frac{e}{\hbar}\right)^3
\begin{tikzpicture}[baseline=(b.center)]
  \begin{feynhand}
    \vertex (a) at (0,0);
    \vertex (b) at (1,0);
    \propag[fermion] (a) to[half left]    node[midway,above] {$i\omega'+i\omega_1+i\omega_2,a$} (b);
    \propag[fermion] (b) to[half left]    node[midway,below] {$i\omega',b$} (a);
    \node[dot] at (a) {};
    \propag[photon] (a) to ++(-1,1) node[left] {$\alpha,i\omega_1$};
    \propag[photon] (a) to ++(-1,-1) node[left] {$\beta,i\omega_2$};
    \node[dot] at (b) {};
    \propag[photon]  (b) to ++(1,0)      node[right] {$\mu,i\omega_1+i\omega_2$};
  \end{feynhand}
\end{tikzpicture}
\;\nonumber\\ 
&-\frac{1}{2}\left(\frac{e}{\hbar}\right)^3
\begin{tikzpicture}[baseline=(b.center)]
  \begin{feynhand}
    \vertex (a) at (0,0);
    \vertex (b) at (1,0);
    \propag[fermion] (a) to[half left]    node[midway,above] {$i\omega'+i\omega_2,a$} (b);
    \propag[fermion] (b) to[half left]    node[midway,below] {$i\omega',b$} (a);
    \node[dot] at (a) {};
    \propag[photon]  (a) to ++(-1,0)      node[left]   {$\beta,i\omega_2$};
    \node[dot] at (b) {};
    \propag[photon]  (b) to ++( 1, 1)     node[right]  {$\mu,i\omega_1+i\omega_2$};
    \propag[photon]  (b) to ++( 1,-1)     node[right]  {$\alpha,i\omega_1$};
  \end{feynhand}
\end{tikzpicture}
\;\nonumber\\ 
&-\frac{1}{2}\left(\frac{e}{\hbar}\right)^3
\begin{tikzpicture}[baseline=(b.center)]
  \begin{feynhand}
    \vertex (a) at (0,0);
    \vertex (b) at (1,0);
    \propag[fermion] (a) to[half left]    node[midway,above] {$i\omega'+i\omega_1,a$} (b);
    \propag[fermion] (b) to[half left]    node[midway,below] {$i\omega',b$} (a);
    \node[dot] at (a) {};
    \propag[photon]  (a) to ++(-1,0)      node[left]   {$\alpha,i\omega_1$};
    \node[dot] at (b) {};
    \propag[photon]  (b) to ++( 1, 1)     node[right]  {$\mu,i\omega_1+i\omega_2$};
    \propag[photon]  (b) to ++( 1,-1)     node[right]  {$\beta,i\omega_2$};
  \end{feynhand}
\end{tikzpicture}
\;\nonumber\\ 
&-\frac{1}{2}\left(\frac{e}{\hbar}\right)^3
\begin{tikzpicture}[baseline=(v2.center)]
  \begin{feynhand}
    \vertex (v1) at (0,-0.5);
    \vertex (v2) at (0.75,0);
    \vertex (v3) at (0,0.5);
    \node[dot]     at (v1) {};
    \node[dot]     at (v2) {};
    \node[dot]     at (v3) {};
    \propag[fermion] (v1) to           node[left]  {$i\omega'+i\omega_1,b$}        (v3);
    \propag[fermion] (v3) to           node[above right] {$i\omega'+i\omega_{1}+i\omega_2,a$} (v2);
    \propag[fermion] (v2) to           node[below right] {$i\omega',c$}        (v1);
    \propag[photon]  (v1) to ++(-1,-0.7) node[left]  {$\alpha,i\omega_1$};
    \propag[photon]  (v3) to ++(-1,0.7) node[left]  {$\beta,i\omega_2$};
    \propag[photon]  (v2) to ++(1, 0)    node[right] {$\mu,i\omega_1+i\omega_2$};
  \end{feynhand}
\end{tikzpicture} 
\;\nonumber\\ 
&-\frac{1}{2}\left(\frac{e}{\hbar}\right)^3
\begin{tikzpicture}[baseline=(v2.center)]
  \begin{feynhand}
    \vertex (v1) at (0,-0.5);
    \vertex (v2) at (0.75,0);
    \vertex (v3) at (0,0.5);
    \node[dot]     at (v1) {};
    \node[dot]     at (v2) {};
    \node[dot]     at (v3) {};
    \propag[fermion] (v1) to           node[left]  {$i\omega'+i\omega_2,b$}        (v3);
    \propag[fermion] (v3) to           node[above right] {$i\omega'+i\omega_{1}+i\omega_2,a$} (v2);
    \propag[fermion] (v2) to           node[below right] {$i\omega',c$}        (v1);
    \propag[photon]  (v3) to ++(-1,0.7) node[left]  {$\alpha,i\omega_1$};
    \propag[photon]  (v1) to ++(-1,-0.7) node[left]  {$\beta,i\omega_2$};
    \propag[photon]  (v2) to ++(1, 0)    node[right] {$\mu,i\omega_1+i\omega_2$};
  \end{feynhand}
\end{tikzpicture} \label{eq:current_response_function_diagram}
\end{align}
We derive $\mathcal{K}^{\mu\alpha\beta}(\omega_1,\omega_2)$ by evaluating the Matsubara frequency sum and performing the analytic continuation of the bosonic frequencies $i\omega_1\to \omega_1,\ i\omega_2\to \omega_2$.
We will show that there is no divergence in $\mathcal{K}^{\mu\alpha\beta}(\omega_1,\omega_2)$ in the static limit $\omega_1,\omega_2 \to 0$.
For notational simplicity, we introduce $\kappa^{\mu\alpha\beta}(x,\omega_1,\omega_2)$ and write
\begin{align}
  &\mathcal{K}^{\mu\alpha\beta}(\omega_1,\omega_2)= \nonumber \\
    &-\frac{1}{2}\left(\frac{e}{\hbar}\right)^3\frac{2i\Gamma}{2\pi i}\int dx f(x) \int\smrm{BZ} \frac{d^d k}{(2\pi)^d}\kappa^{\mu\alpha\beta}(x,\omega_1,\omega_2). \label{eq:current_response_function}
\end{align}

\subsubsection{$O(\omega^0)$ terms}
Setting $\omega_1=\omega_2=0$ in $\kappa^{\mu\alpha\beta}(x,\omega_1,\omega_2)$, we obtain 
\begin{align}
  \kappa^{\mu\alpha\beta} (x,0,0)
  =& \partial_{k_\beta} (\mathrm{Tr}[H^{\mu\alpha}G^R G^A]) \label{eq:0linTad} \\
  &+\partial_{k_\beta} (\mathrm{Tr}[H^{\mu}G^R H^{\alpha} G^R G^A]) \label{eq:0linbub1} \\
  &+\partial_{k_\beta} (\mathrm{Tr}[H^{\mu}G^R G^A H^{\alpha} G^A]). \label{eq:0linbub2}
\end{align}
Thus, $\mathcal{K}^{\mu\alpha\beta}(0,0)$ vanishes upon integration over the Brillouin zone, since the integrand $\kappa^{\mu\alpha\beta}(x,0,0)$ is a total derivative with respect to $k_\beta$.
The terms in Eqs.~\eqref{eq:0linTad}, \eqref{eq:0linbub1}, and \eqref{eq:0linbub2} correspond, respectively, to the $k_\beta$-derivatives of the tadpole and bubble diagram contributions appearing in the response coefficient $\mathcal{K}^{\mu\alpha}$ of the linear response $j_\mu=\mathcal{K}^{\mu\alpha}A_\alpha$.
In other words, they describe the change of this linear-response coefficient in response to $A_\beta$. 
Since a uniform static vector potential $A_\beta$ does not affect physical observables, the corresponding Brillouin-zone integral must vanish. 
Here, we have derived the total-derivative form with respect to $k_\beta$, but the same procedure can also be applied with respect to $k_\alpha$.

\subsubsection{$O(\omega^1)$ terms}
A similar total-derivative argument can be applied to the $O(\omega)$ terms although some diagrams drop out because $\partial_{\omega_{1(2)}}$ acts on $\kappa^{\mu\alpha\beta}(\omega_1,\omega_2)$ before setting $\omega_1=\omega_2=0$.
In addition, due to the frequency derivative $\partial_{\omega_{1(2)}}$, $\partial_{\omega_{1(2)}}\kappa^{\mu\alpha\beta} (x,\omega_1,\omega_2)|_{\omega_1=0,\omega_2=0}$ reduces to a total derivative only with respect to $k_{\beta(\alpha)}$, whereas $\kappa^{\mu\alpha\beta} (x,0,0)$ reduces to a total derivative with respect to either $k_\alpha$ or $k_\beta$.

Indeed, for $\partial_{\omega_1}$, we obtain
\begin{align}
  \partial_{\omega_1}\kappa (x,\omega_1,\omega_2)|_{\omega_1=0,\omega_2=0}
  =&-\hbar\partial_{k_\beta} (\mathrm{Tr}[H^{\mu}[G^R]^2 H^{\alpha} G^R G^A]) \label{eq:1linbub1} \\
  &+\hbar\partial_{k_\beta} (\mathrm{Tr}[H^{\mu}G^R G^A H^{\alpha} [G^A]^2]) \label{eq:1linbub2}
\end{align}
which is simply obtained by taking $\partial_{\omega_1}$ of the finite-frequency linear-response functions corresponding to \eqref{eq:0linTad}, \eqref{eq:0linbub1}, and \eqref{eq:0linbub2}.

Physically, these terms represent how the linear response to a static $E_\alpha$ changes in the presence of a static $A_\beta$, and thus $\partial_{\omega_{1}}\mathcal{K}^{\mu\alpha\beta}(\omega_1,\omega_2)|_{\omega_1=0,\omega_2=0}$ vanishes because a static vector potential does not produce any physical effect.

\subsubsection{$O(\omega^2)$ terms}
Similarly, for $\partial_{\omega_1}^2$, we obtain
\begin{align}
  \partial_{\omega_1}^2\kappa (x,\omega_1,\omega_2)|_{\omega_1=0,\omega_2=0}
  =&\hbar^2\partial_{k_\beta} (\mathrm{Tr}[H^{\mu} 2[G^R]^3 H^{\alpha} G^R G^A]) \label{eq:2linbub1} \\
  &+\hbar^2\partial_{k_\beta} (\mathrm{Tr}[H^{\mu}G^R G^A H^{\alpha} 2[G^A]^3]) \label{eq:2linbub2}.
\end{align}

The physical interpretation of these terms is also similar to that of Eqs.~\eqref{eq:1linbub1} and \eqref{eq:1linbub2}.
These terms represent how the $O(\omega_1)$ contribution to the response function of $j_\mu$ induced by $E_\alpha(\omega_1)$ changes in the presence of a static $A_\beta$.
Therefore, $\partial_{\omega_{1}}^2\mathcal{K}^{\mu\alpha\beta}(\omega_1,\omega_2)|_{\omega_1=0,\omega_2=0}$ vanishes because a static vector potential does not produce any physical effect.

By contrast, $\partial_{\omega_1}\partial_{\omega_2}\kappa (x,\omega_1,\omega_2)|_{\omega_1=0,\omega_2=0}$ indeed contributes to the physical response because this term does not reduce to a total derivative with respect to either $k_\alpha$ or $k_\beta$.

\section{Derivation of injection current and shift current} \label{app:injectionShift}
In this section, we evaluate each diagram in the current response function (Eq.~\eqref{eq:current_response_function_diagram}).
First, we evaluate the first and second diagrams in Eq.~\eqref{eq:current_response_function_diagram} as
\begin{align}
  &\begin{tikzpicture}[baseline=(b.center)]
  \begin{feynhand}
    \vertex (a) at (0,0);
    \vertex (b) at (1,0);
    \propag[fermion] (a) to[half left]    node[midway,above] {$i\omega',a$} (b);
    \propag[plain] (b) to[half left] (a);
    \node[dot] at (b) {};
    \propag[photon]  (b) to ++(1, 1)     node[left]  {$\alpha,i\omega_1$};
    \propag[photon]  (b) to ++(1,-1)     node[left]  {$\beta,i\omega_2$};
    \propag[photon]  (b) to ++( 1, 0)     node[right] {$\mu,i\omega_1+i\omega_2$};
  \end{feynhand}
\end{tikzpicture} \nonumber\\
+&\begin{tikzpicture}[baseline=(b.center)]
  \begin{feynhand}
    \vertex (a) at (0,0);
    \vertex (b) at (1,0);
    \propag[fermion] (a) to[half left]    node[midway,above] {$i\omega'+i\omega_1+i\omega_2,a$} (b);
    \propag[fermion] (b) to[half left]    node[midway,below] {$i\omega',b$} (a);
    \node[dot] at (a) {};
    \propag[photon] (a) to ++(-1,1) node[left] {$\alpha,i\omega_1$};
    \propag[photon] (a) to ++(-1,-1) node[left] {$\beta,i\omega_2$};
    \node[dot] at (b) {};
    \propag[photon]  (b) to ++(1,0)      node[right] {$\mu,i\omega_1+i\omega_2$};
  \end{feynhand}
\end{tikzpicture}\nonumber\\
\to &\frac{2i\Gamma}{2\pi i} \int dx f(x) \mathrm{tr}[H^{\mu\alpha\beta} G^A(x)G^R(x)] \nonumber \\
+&\frac{2i\Gamma}{2\pi i} \int dx f(x) \mathrm{tr}[H^{\mu}G^R(x)H^{\alpha\beta} G^A(x)G^R(x)] \nonumber \\
+&\frac{2i\Gamma}{2\pi i} \int dx f(x) \mathrm{tr}[H^{\mu}G^A(x)G^R(x)H^{\alpha\beta} G^A(x)] \nonumber \\
=&\frac{2i\Gamma}{2\pi i} \int dx f(x) \int\smrm{BZ} \frac{d^d k}{(2\pi)^d}\partial_{k_\mu} \mathrm{Tr}[H^{\alpha\beta} G^A(x)G^R(x)], \label{eq:tadpoleDerivative}
\end{align}
where $\mathrm{tr}[\cdots]\equiv \int\smrm{BZ} \frac{d^d k}{(2\pi)^d}\mathrm{Tr}[\cdots]$ and $\to$ indicates the analytic continuation $i\omega_1\to \omega_1,\ i\omega_2\to \omega_2$.
Here, we use the relation $-2i\Gamma G^R(x)G^A(x)=(G^R(x)-G^A(x))$.
These terms correspond to the $k_\mu$-derivative of the tadpole diagram contribution appearing in the response coefficient $\sigma^{\alpha\beta}$ of the linear response $j_\alpha=\sigma^{\alpha\beta}E_\beta$,
and thus they reduce to the total derivative with respect to $k_\mu$.

The third diagram in Eq.~\eqref{eq:current_response_function_diagram} is evaluated as
\begin{align}  
&\begin{tikzpicture}[baseline=(b.center)]
  \begin{feynhand}
    \vertex (a) at (0,0);
    \vertex (b) at (1,0);
    \propag[fermion] (a) to[half left]    node[midway,above] {$i\omega'+i\omega_2,a$} (b);
    \propag[fermion] (b) to[half left]    node[midway,below] {$i\omega',b$} (a);
    \node[dot] at (a) {};
    \propag[photon]  (a) to ++(-1,0)      node[left]   {$\beta,i\omega_2$};
    \node[dot] at (b) {};
    \propag[photon]  (b) to ++( 1, 1)     node[right]  {$\mu,i\omega_1+i\omega_2$};
    \propag[photon]  (b) to ++( 1,-1)     node[right]  {$\alpha,i\omega_1$};
  \end{feynhand}
\end{tikzpicture}
\;\nonumber\\ 
\to &\frac{2i\Gamma}{2\pi i} \int dx f(x) \mathrm{tr}[H^{\mu\alpha}G^R(x-\hbar\omega)H^\beta G^R(x)G^A(x)]\nonumber \\
+&\frac{2i\Gamma}{2\pi i} \int dx f(x) \mathrm{tr}[H^{\mu\alpha}G^R(x)G^A(x)H^\beta G^A(x+\hbar\omega)] \nonumber \\
=&\frac{2i\Gamma}{2\pi i} \int dx f(x) \mathrm{tr}[H^{\mu\alpha}G^A(x-\hbar\omega)(1-2i\Gamma G^R(x-\hbar \omega))\nonumber \\
&\qquad \times H^\beta G^R(x)G^A(x)] \label{eq:m2}\\
+&\frac{2i\Gamma}{2\pi i} \int dx f(x) \mathrm{tr}[H^{\mu\alpha}G^R(x)G^A(x)H^\beta\nonumber \\
&\qquad \times  G^R(x+\hbar\omega) (1+2i\Gamma G^A(x+\hbar\omega))]. \label{eq:p4}
\end{align}
Here, we use $G^R(x)=G^A(x)(1-2i\Gamma G^R(x))$ and $G^A(x)=(1+2i\Gamma G^A(x))G^R(x)$.
Note that $[G^R(x),G^A(y)]=0$, $[G^R(x),G^R(y)]=0$, and $[G^A(x),G^A(y)]=0$ for any $x,y$.
In the following, we will use this relation to evaluate the diagrams.
The fourth diagram in Eq.~\eqref{eq:current_response_function_diagram} is evaluated as
\begin{align}
&\begin{tikzpicture}[baseline=(b.center)]
  \begin{feynhand}
    \vertex (a) at (0,0);
    \vertex (b) at (1,0);
    \propag[fermion] (a) to[half left]    node[midway,above] {$i\omega'+i\omega_1,a$} (b);
    \propag[fermion] (b) to[half left]    node[midway,below] {$i\omega',b$} (a);
    \node[dot] at (a) {};
    \propag[photon]  (a) to ++(-1,0)      node[left]   {$\alpha,i\omega_1$};
    \node[dot] at (b) {};
    \propag[photon]  (b) to ++( 1, 1)     node[right]  {$\mu,i\omega_1+i\omega_2$};
    \propag[photon]  (b) to ++( 1,-1)     node[right]  {$\beta,i\omega_2$};
  \end{feynhand}
\end{tikzpicture}
\;\nonumber\\ 
\to &\frac{2i\Gamma}{2\pi i} \int dx f(x) \mathrm{tr}[H^{\mu\beta}G^R(x+\hbar\omega)H^\alpha G^R(x)G^A(x)]\label{eq:p2}\\
+&\frac{2i\Gamma}{2\pi i} \int dx f(x) \mathrm{tr}[H^{\mu\beta}G^R(x)G^A(x)H^\alpha G^A(x-\hbar\omega)]. \label{eq:m4}
\end{align}
The fifth diagram in Eq.~\eqref{eq:current_response_function_diagram} is evaluated as
\begin{align}
&\begin{tikzpicture}[baseline=(v2.center)]
  \begin{feynhand}
    \vertex (v1) at (0,-0.5);
    \vertex (v2) at (0.75,0);
    \vertex (v3) at (0,0.5);
    \node[dot]     at (v1) {};
    \node[dot]     at (v2) {};
    \node[dot]     at (v3) {};
    \propag[fermion] (v1) to           node[left]  {$i\omega'+i\omega_1,b$}        (v3);
    \propag[fermion] (v3) to           node[above right] {$i\omega'+i\omega_{1}+i\omega_2,a$} (v2);
    \propag[fermion] (v2) to           node[below right] {$i\omega',c$}        (v1);
    \propag[photon]  (v1) to ++(-1,-0.7) node[left]  {$\alpha,i\omega_1$};
    \propag[photon]  (v3) to ++(-1,0.7) node[left]  {$\beta,i\omega_2$};
    \propag[photon]  (v2) to ++(1, 0)    node[right] {$\mu,i\omega_1+i\omega_2$};
  \end{feynhand}
\end{tikzpicture} 
\;\nonumber\\ 
\to & \frac{2i\Gamma}{2\pi i} \int dx f(x) \mathrm{tr}[H^{\mu}G^R(x)H^\beta G^R(x+\hbar\omega )H^\alpha G^A(x)G^R(x)]\nonumber \\
+&\frac{2i\Gamma}{2\pi i} \int dx f(x) \mathrm{tr}[H^{\mu}G^R(x-\hbar\omega)H^\beta G^R(x)G^A(x)H^\alpha G^A(x-\hbar\omega)]\nonumber \\
+&\frac{2i\Gamma}{2\pi i} \int dx f(x) \mathrm{tr}[H^{\mu}G^A(x)G^R(x)H^\beta G^A(x+\hbar\omega)H^\alpha G^A(x)] \nonumber \\
= & \frac{2i\Gamma}{2\pi i} \int dx f(x) \mathrm{tr}[H^{\mu}G^R(x)H^\beta G^R(x+\hbar\omega )H^\alpha G^A(x)G^R(x)] \label{eq:p1}\\
+&\frac{2i\Gamma}{2\pi i} \int dx f(x) \mathrm{tr}[H^{\mu}G^A(x-\hbar\omega)(1-2i\Gamma G^R(x-\hbar\omega))\nonumber \\
&\qquad \times H^\beta G^R(x)G^A(x)H^\alpha G^A(x-\hbar\omega)] \label{eq:m3}\\
+&\frac{2i\Gamma}{2\pi i} \int dx f(x) \mathrm{tr}[H^{\mu}G^A(x)G^R(x)H^\beta \nonumber \\
&\qquad \times G^R(x+\hbar\omega)(1+2i\Gamma G^A(x+\hbar\omega)) H^\alpha G^A(x)]. \label{eq:p5}
\end{align}
The last diagram in Eq.~\eqref{eq:current_response_function_diagram} is evaluated as
\begin{align}
&\begin{tikzpicture}[baseline=(v2.center)]
  \begin{feynhand}
    \vertex (v1) at (0,-0.5);
    \vertex (v2) at (0.75,0);
    \vertex (v3) at (0,0.5);
    \node[dot]     at (v1) {};
    \node[dot]     at (v2) {};
    \node[dot]     at (v3) {};
    \propag[fermion] (v1) to           node[left]  {$i\omega'+i\omega_2,b$}        (v3);
    \propag[fermion] (v3) to           node[above right] {$i\omega'+i\omega_{1}+i\omega_2,a$} (v2);
    \propag[fermion] (v2) to           node[below right] {$i\omega',c$}        (v1);
    \propag[photon]  (v3) to ++(-1,0.7) node[left]  {$\alpha,i\omega_1$};
    \propag[photon]  (v1) to ++(-1,-0.7) node[left]  {$\beta,i\omega_2$};
    \propag[photon]  (v2) to ++(1, 0)    node[right] {$\mu,i\omega_1+i\omega_2$};
  \end{feynhand}
\end{tikzpicture}\nonumber \\
\to & \frac{2i\Gamma}{2\pi i} \int dx f(x) \mathrm{tr}[H^{\mu}G^R(x)H^\alpha G^R(x-\hbar\omega )H^\beta G^A(x)G^R(x)]\nonumber \\
+&\frac{2i\Gamma}{2\pi i} \int dx f(x) \mathrm{tr}[H^{\mu}G^R(x+\hbar\omega)H^\alpha G^R(x)G^A(x)H^\beta G^A(x+\hbar\omega)]\nonumber \\
+&\frac{2i\Gamma}{2\pi i} \int dx f(x) \mathrm{tr}[H^{\mu}G^A(x)G^R(x)H^\alpha G^A(x-\hbar\omega)H^\beta G^A(x)] \nonumber \\
= & \frac{2i\Gamma}{2\pi i} \int dx f(x) \mathrm{tr}[H^{\mu}G^R(x)H^\alpha G^A(x-\hbar\omega )(1-2i\Gamma G^R(x-\hbar\omega)) \nonumber \\
&\qquad \times H^\beta G^A(x)G^R(x)] \label{eq:m1}\\
+&\frac{2i\Gamma}{2\pi i} \int dx f(x) \mathrm{tr}[H^{\mu}G^R(x+\hbar\omega)H^\alpha G^R(x)G^A(x)H^\beta \nonumber \\
&\qquad \times (1+2i\Gamma G^A(x+\hbar\omega)) G^R(x+\hbar\omega)] \label{eq:p3}\\
+&\frac{2i\Gamma}{2\pi i} \int dx f(x) \mathrm{tr}[H^{\mu}G^A(x)G^R(x)H^\alpha G^A(x-\hbar\omega)H^\beta G^A(x)]. \label{eq:m5}
\end{align}
Collecting the terms in Eqs.~\eqref{eq:m4}, \eqref{eq:m5}, \eqref{eq:m1}, \eqref{eq:m2}, and \eqref{eq:m3} that are explicitly linear in $\Gamma$ yields
\begin{align}
  \frac{2i\Gamma}{2\pi i} \int dx f(x)\int\smrm{BZ} \frac{d^d k}{(2\pi)^d} \partial_{k_\mu}\Tr [ H^\beta G^A(x) G^R(x)H^\alpha G^A(x-\hbar\omega)]. \label{eq:m1m5}
\end{align}
Here, Eqs.~\eqref{eq:m4}, \eqref{eq:m5}, \eqref{eq:m1}, \eqref{eq:m2}, and \eqref{eq:m3} correspond to the derivatives acting on each factor of the above expression in this order.
Similarly, collecting the terms in Eqs.~\eqref{eq:p3}, \eqref{eq:p4}, \eqref{eq:p5}, \eqref{eq:p1}, and \eqref{eq:p2} that are explicitly linear in $\Gamma$ yields
\begin{align}
  \frac{2i\Gamma}{2\pi i} \int dx f(x) \int\smrm{BZ} \frac{d^d k}{(2\pi)^d}\partial_{k_\mu} \Tr[H^\beta G^R(x+\hbar\omega) H^\alpha G^A(x)G^R(x)]. \label{eq:p1p5}
\end{align}
In this case, Eqs.~\eqref{eq:p3}, \eqref{eq:p4}, \eqref{eq:p5}, \eqref{eq:p1}, and \eqref{eq:p2} represent the derivatives acting on each factor of the above expression in this order.
Therefore, Eqs.~\eqref{eq:m1m5} and \eqref{eq:p1p5} are the $k_\mu$-derivatives of the bubble diagrams of the linear response function $\sigma^{\alpha\beta}$.
Thus, Eqs.~\eqref{eq:tadpoleDerivative}, \eqref{eq:m1m5}, and \eqref{eq:p1p5} are total derivatives with respect to $k_\mu$ of the linear response function $\sigma^{\alpha\beta}$, which vanish in $\mathcal{K}^{\mu\alpha\beta}(\omega,-\omega)$
after integrating over the Brillouin zone.

Thus, the only remaining terms are terms that explicitly contain $\Gamma^2$.
Collecting Eqs.~\eqref{eq:m3}, \eqref{eq:m2}, and \eqref{eq:m1} gives
\begin{align}
  -\frac{1}{2\pi i} \int dx 4\Gamma^2 (-f(x)) \mathrm{tr} [&\partial_{k_\mu} (G^R(x) H^\alpha G^A(x-\hbar\omega) )\nonumber \\
  &\times G^R(x-\hbar\omega) H^\beta G^A(x) ]
\end{align}
Similarly, collecting Eqs.~\eqref{eq:p3}, \eqref{eq:p4}, and \eqref{eq:p5} gives
\begin{align}
  -\frac{1}{2\pi i} \int dx 4\Gamma^2 f(x) \mathrm{tr}[&\partial_{k_\mu} (G^R(x+\hbar\omega) H^\alpha G^A(x) )\nonumber \\
  &\times G^R(x) H^\beta G^A(x+\hbar\omega) ]
\end{align}
Therefore, we obtain Eq.~\eqref{eq:currentbyExcitation}
\begin{align}
  \mathcal{K}&^{\mu\alpha\beta}(\omega,-\omega)=\left(\frac{e}{\hbar}\right)^3\int \frac{dx}{2\pi i}\int \frac{d^d k}{(2\pi)^d} 2\Gamma^2 (f(x)-f(x+\hbar\omega)) \nonumber \\
  &\times \Tr[\partial_{k_\mu} (G^R(x+\hbar\omega) H^\alpha G^A(x) )G^R(x) H^\beta G^A(x+\hbar\omega) ] \label{eq:currentbyExcitationInX}
\end{align}
Using $G^R(x) H^\alpha G^A(y)=\sum_{ab}\ket{u_a}\bra{u_a}H^\alpha\ket{u_b}\bra{u_b}/((x-\xi_a+i\Gamma)(y-\xi_b-i\Gamma))$, the above expression can be rewritten as
\begin{align}
  &\mathcal{K}^{\mu\alpha\beta}(\omega,-\omega)
  \nonumber \\
  =&\left(\frac{e}{\hbar}\right)^3\int \frac{dx}{2\pi i}\int \frac{d^d k}{(2\pi)^d} 2\Gamma^2 (f(x)-f(x+\hbar\omega)) \nonumber \\
  &\times \Bigg\{\sum_{ab}\frac{1}{[(x+\hbar\omega-\xi_a)^2+\Gamma^2][(x-\xi_b)^2+\Gamma^2]} \nonumber \\
  &\times \Bigg[\Tr [\partial_\mu (\ket{u_a}\bra{u_a}H^\alpha\ket{u_b}\bra{u_b})\ket{u_b}\bra{u_b}H^\beta\ket{u_a}\bra{u_a}] \nonumber \\
  &+H_{ab}^\alpha H_{ba}^\beta \left(\frac{\xi^\mu_a}{x+\hbar\omega-\xi_a+i\Gamma}+\frac{\xi^\mu_b}{x-\xi_b-i\Gamma}\right)\Bigg]
  \Bigg\}+O(\Gamma)
\end{align}
Here, we assume that the bands are nondegenerate. 
Under this assumption, the terms in the third line with $b\neq c$ or $a\neq d$ are $O(\Gamma)$ and therefore vanish in the clean limit $\Gamma \to 0$.
By using the Taylor's theorem in terms of $F(x)\equiv f(x)-f(x+\hbar\omega)$, we obtain
\begin{align}
  F(x)=&F(\xi_a-\hbar \omega)+F'(\xi_a-\hbar \omega)(x-\xi_a+\hbar\omega)+r_2^a(x) \\
  =&F(\xi_a-\hbar \omega)+F'(\xi_a-\hbar \omega)(x-\xi_a+\hbar\omega)\nonumber \\
  &+\frac{1}{2}F''(\xi_a-\hbar \omega)(x-\xi_a+\hbar\omega)^2+r_3^a(x) .
\end{align}
For the Fermi--Dirac distribution function, we have $f'(x)=-\beta f(x)[1-f(x)]$.
Since $f^{(n)}(x)$ can be expressed as a polynomial in $f(x)$ and $0\leq f(x)\leq 1$, all derivatives $f^{(n)}(x)$ are bounded.
It follows that $F^{(n)}(x)$ is also bounded, since $F(x)=f(x)-f(x+\hbar\omega)$.
Consequently, there exists a positive constant $M_n$ such that the Taylor remainder satisfies $|r_n^a(x)|\leq M_n|x-\xi_a+\hbar\omega|^n$.
Therefore, we obtain
\begin{align}
  &\left|\Gamma^2\int dx\frac{r_2^a(x)}{[(x+\hbar\omega-\xi_a)^2+\Gamma^2][(x-\xi_b)^2+\Gamma^2]}\right|\nonumber\\
  &\leq M_2\Gamma^2\int dx\frac{(x-\xi_a+\hbar\omega)^2}{[(x+\hbar\omega-\xi_a)^2+\Gamma^2][(x-\xi_b)^2+\Gamma^2]}\nonumber\\
  &\leq\pi M_2\Gamma=O(\Gamma),\\
  &\left|\Gamma^2\int \frac{r_3^a(x)dx}{(x+\hbar\omega-\xi_a+i\Gamma)^2(x+\hbar\omega-\xi_a-i\Gamma)[(x-\xi_b)^2+\Gamma^2]}\right|\nonumber\\
  &\leq M_3\Gamma^2\int dx\frac{|x-\xi_a+\hbar\omega|^3}{[(x+\hbar\omega-\xi_a)^2+\Gamma^2]^{3/2}[(x-\xi_b)^2+\Gamma^2]}\nonumber\\
  &\leq \pi M_3\Gamma=O(\Gamma).
\end{align}
By performing the integral in terms of $x$ with the residue theorem, we obtain
\begin{widetext}
\begin{align}
  \mathcal{K}^{\mu\alpha\beta}(\omega,-\omega)=
  &\left(\frac{e}{\hbar}\right)^3\int \frac{d^d k}{(2\pi)^d} \Bigg\{\sum_{a\neq b}\Tr [\partial_\mu (\ket{u_a}\mel{u_a}{\partial_\alpha}{u_b} \bra{u_b}\xi_{ba})\ket{u_b}\mel{u_b}{\partial_\beta}{u_a} \bra{u_a}\xi_{ab}]\frac{-iF(\xi_a-\hbar\omega)2\Gamma}{(\xi_{ba}+\hbar\omega)^2+4\Gamma^2} \nonumber \\
  &+\sum_{a\neq b}H_{ab}^\alpha H_{ba}^\beta (\xi^\mu_aF(\xi_a-\hbar\omega)-\xi^\mu_bF(\xi_b))\left[-\frac{1}{(\xi_{ba}+\hbar\omega)^2+4\Gamma^2} +\frac{1}{2(\xi_{ba}+\hbar\omega+2i\Gamma)^2}\right]\nonumber \\
  &+\sum_{a\neq b}H_{ab}^\alpha H_{ba}^\beta [\xi^\mu_aF'(\xi_a-\hbar\omega)+\xi^\mu_b F'(\xi_b)] i\Gamma\left[-\frac{1}{(\xi_{ba}+\hbar\omega)^2+4\Gamma^2} -\frac{1}{2(\xi_{ba}+\hbar\omega+2i\Gamma)^2}\right]\nonumber \\
  &-\frac{\hbar\omega+6i\Gamma}{2(\hbar\omega+2i\Gamma)[(\hbar\omega)^2+4\Gamma^2]}\sum_{a}\xi_a^\alpha \xi_a^\beta \xi^\mu_a\big[F(\xi_a-\hbar\omega)-F(\xi_a)\big]\Bigg\} +O(\Gamma)
\end{align}
\end{widetext}
Assuming a small $\Gamma$ compared to the band geometry and band dispersion, we can ignore the $(\xi_{ba}+\hbar\omega+2i\Gamma)^{-2}$ terms because $\int d\xi_a (\xi_{ba}+\hbar\omega+2i\Gamma)^{-2}=0$.
The first term in the braket of the third line vanishes when the excitation points are far from the Fermi surface, since $F'(\xi_a-\hbar\omega)=f'(\xi_a-\hbar\omega)-f'(\xi_a)$.
By using the relation $\frac{2\Gamma}{(\xi_{ba}+\hbar\omega)^2+4\Gamma^2}\simeq \pi \delta(\xi_{ba}+\hbar\omega)$ and $\frac{1}{(\xi_{ba}+\hbar\omega)^2+4\Gamma^2}\simeq \pi \delta(\xi_{ba}+\hbar\omega)\tau /\hbar$, we obtain
\begin{align}
 \mathcal{K}^{\mu\alpha\beta}(\omega,-\omega)\simeq &\omega^2\frac{e^3\pi}{\hbar}\int \frac{d^d k}{(2\pi)^d} \sum_{ab}f_{ba}\delta(\hbar\omega+\Delta_{ba}) \nonumber \\
  &\times \left[Q_{ba}^{\alpha\beta}\left(\frac{\tau \partial_{k_\mu} \Delta_{ab}}{\hbar}+\frac{i\partial_{k_\mu} \Delta_{ab}}{\Delta_{ab}}\right)+iC_{ab}^{\mu\beta\alpha}\right]\nonumber \\
  &-\frac{e^3}{2\hbar^5\omega^2}\int \frac{d^d k}{(2\pi)^d} \sum_{a}\xi_a^\alpha \xi_a^\beta \xi^\mu_a\nonumber \\
  &\times \big[f(\xi_a-\hbar\omega)-2f(\xi_a)+f(\xi_a+\hbar\omega)\big]
  \label{eq:halfPhotocurrent}
\end{align}
where $Q_{ba}^{\alpha\beta}\equiv\mathrm{Tr}[P_a \partial_\alpha P_b \partial_\beta P_a]$ is the two-state quantum geometric tensor, and the two-state quantum geometric connection $C_{ab}^{\mu\beta\alpha}\equiv\mathrm{Tr}[P_b\partial_\beta P_a (\partial_\mu \partial_\alpha P_b +\partial_\mu P_a \partial_\alpha P_b)]$ where $P_a\equiv\ket{u_a}\bra{u_a}$ is the projection operator onto the $a$-th band~\cite{Ahn2022,Mitscherling2025}.
The last term corresponds to the intraband contribution to current
rectification discussed in Ref.~\cite{Matsyshyn2023}.
As can be seen from \eqref{eq:DCsumOfalpha_beta}, $\mathcal{K}^{\mu\alpha\beta}(\omega,-\omega)+\mathcal{K}^{\mu\beta\alpha}(-\omega,\omega)$ yields the injection current and the shift current in the clean limit $\Gamma \to 0$.
Note that $\frac{i\partial_{k_\mu} \Delta_{ab}}{\Delta_{ab}}$ term cancels out in $\mathcal{K}^{\mu\alpha\beta}(\omega,-\omega)+\mathcal{K}^{\mu\beta\alpha}(-\omega,\omega)$.

\section{Derivation of the Drude term, the Berry curvature dipole term and the quantum metric dipole term} \label{app:DrudeBCD}
Using Eqs.~\eqref{eq:nonReciprocalEsquared} and \eqref{eq:currentbyExcitationInX}, we obtain the nonlinear response function induced by a static electric field as
\begin{align}
  &\sigma^{\mu\alpha\beta}\nonumber \\
  \equiv &\frac{1}{2}\partial_\omega^2\mathcal{K}^{\mu\alpha\beta}(\omega,-\omega)|_{\omega=0}\nonumber \\
  =&-\frac{e^3}{\hbar}\frac{\Gamma^2}{2\pi i}\int dx \int\smrm{BZ} \frac{d^d k}{(2\pi)^d} f'(x) \nonumber \\
  &\times \Tr[\partial_{k_\mu} (G^R H^\beta G^A) G^R (G^R H^\alpha -H^\alpha G^A) G^A +(\alpha \leftrightarrow \beta)].\label{eq:sigmaMuAlphaBetaGreenFunction}
\end{align}
Here, we write $G^R=G^R(x)$ and $G^A=G^A(x)$ for brevity.

In what follows, we evaluate the integrals in the above expression,
\begin{align}
  &\sigma^{\mu\alpha\beta}\nonumber \\
  =&-\frac{e^3}{\hbar}\frac{\Gamma^2 }{2\pi i} \int dx \int \frac{d^d k}{(2\pi)^d}f'(x)\nonumber \\
  &\times \Big[\sum_{abc}H_{bc}^\mu H_{ca}^\beta H_{ab}^\alpha G^R_c \left(G^R_a-G^A_b\right)G^R_a G^A_a G^R_b G^A_b  \nonumber \\
  &+\sum_{ab}H_{ba}^{\mu\beta} H_{ab}^\alpha\left(G^R_a-G^A_b\right)G^R_a G^A_a G^R_b G^A_b \nonumber \\
  &+\sum_{abc}H_{ca}^\mu H_{ab}^\alpha H_{bc}^\beta G^A_c \left(G^R_a-G^A_b\right)G^R_a G^A_a G^R_b G^A_b +(\alpha \leftrightarrow \beta)\Big] \nonumber \\
  =&\frac{e^3}{\hbar}\int \frac{d^d k}{(2\pi)^d}\sum_{ab}\Bigg\{ \Big[\sum_{c}H_{bc}^\mu H_{ca}^\beta H_{ab}^\alpha (I_{abc}^{(1)}+I_{abc}^{(2)})  \nonumber \\
  &+H_{ba}^{\mu\beta} H_{ab}^\alpha I_{ab}^{(3)}\Big]+(c.c.,\ (a\leftrightarrow b))\Bigg\}+(\alpha \leftrightarrow \beta), \label{eq:generalNonReciprocalKernel}
\end{align}
with
\begin{align}
  I_{abc}^{(1)}=&-\frac{\Gamma^2 }{2\pi i} \int dx f'(x) G^R_c G^{R2}_a G^A_a G^R_b G^A_b, \\
  I_{abc}^{(2)}=&\frac{\Gamma^2 }{2\pi i} \int dx f'(x) G^R_c G^R_a G^A_a G^R_b G^{A2}_b, \\
  I_{ab}^{(3)}=&-\frac{\Gamma^2 }{2\pi i} \int dx f'(x) G^R_aG^R_a G^A_a G^R_b G^A_b.
\end{align}
Here, $H_{ab}^\alpha=\bra{u_a}\partial_{k_\alpha} H\ket{u_b}$, $H_{ab}^{\alpha\beta}=\bra{u_a}\partial_{k_\alpha} \partial_{k_\beta} H\ket{u_b}$, $G^R_a=\frac{1}{x-\xi_a+i\Gamma},G^A_a=\frac{1}{x-\xi_a-i\Gamma}$.

\subsection{Integrals}
We now consider the clean limit $\Gamma \to 0$ and organize the expression in powers of $\Gamma$.
In the following, we assume that there are no degeneracies.
We divide the analysis into five cases: (1) $a\neq b,\ b\neq c,\ c\neq a$, (2) $a=b\neq c$, (3) $b=c\neq a$, (4) $c=a\neq b$, and (5) $a=b=c$ because the order of the poles in the integrand changes depending on the coincidences among the indices $a$, $b$, and $c$.

\subsubsection{Case $a\neq b,\ b\neq c,\ c\neq a$}
First, we consider the case where $a,b,c$ are all different.
Since all band indices are distinct, no additional pole merging occurs. Expanding the resulting residues in powers of $\Gamma$ gives
\begin{align}
  I_{abc}^{(1)}
&=\Gamma^2\left[
- f_+'(\xi_a+i\Gamma)\frac{1}{(2i\Gamma)^2(\xi_{ab}+2i\Gamma)(\xi_{ac}+2i\Gamma)\xi_{ab}} \right. \nonumber \\
&\left. \hspace{1cm}- f'_{-a}\frac{1}{\xi_{ab}\xi_{ac}(-2i\Gamma)^2(\xi_{ab}-2i\Gamma)}
\right]+O(\Gamma) \nonumber \\
&=f'_a\frac{1}{4\xi_{ab}^2\xi_{ac}}+O(\Gamma),\\
  I_{abc}^{(2)}
&=\Gamma^2\left[
- f_+'(\xi_b+i\Gamma)\frac{1}{(\xi_{ba}+2i\Gamma)(2i\Gamma)^2(\xi_{bc}+2i\Gamma)\xi_{ba}} \right. \nonumber \\
&\left.\hspace{1cm}- f_-'(\xi_b-i\Gamma)\frac{1}{\xi_{ba}\xi_{bc}(\xi_{ba}-2i\Gamma)(-2i\Gamma)^2}
\right] +O(\Gamma) \nonumber \\
&=f'_b\frac{1}{4\xi_{ba}^2\xi_{bc}}+O(\Gamma),\\
  I_{ab}^{(3)}
&= -\Gamma^2\left[f_+'(\xi_a+i\Gamma)\frac{1}{(2i\Gamma)^2\xi_{ab}^2} \right. \nonumber \\
&\left.\hspace{1cm} +f_-'(\xi_a-i\Gamma)\frac{1}{\xi_{ab}^2(-2i\Gamma)^2}\right] +O(\Gamma) \nonumber \\
&= f'_a\frac{1}{4\xi_{ab}^2} +O(\Gamma).
\end{align}
Here, we divide the Fermi distribution function into two parts, $f(x)=f_+(x)+f_-(x)$, where $f_{\pm}(x)\equiv \frac{1}{4}\pm \frac{1}{2\pi i}\psi\left(\frac{1}{2}\pm \frac{\beta x}{2\pi i}\right)$ with $\psi(z)$ being the digamma function.
By construction, $f_+(x)$ and $f_-(x)$ are analytic in the upper half-plane and the lower half-plane, respectively.

\subsubsection{Case $a=b\neq c$}
When $a=b$, the poles at $\xi_b$ merge and higher-order residues must be evaluated.
Expanding these residues up to $O(\Gamma^0)$ gives
\begin{align}
&I_{bbc}^{(1)}\nonumber \\
&=
-\Gamma^2\Biggl\{
f_+''(\xi_b+i\Gamma)\frac{1}{(2i\Gamma)^3}\frac{1}{\xi_{bc}+2i\Gamma} \nonumber \\
&\quad+f_+'(\xi_b+i\Gamma)\left[
-\frac{3}{(2i\Gamma)^4}\frac{1}{\xi_{bc}+2i\Gamma}
+\frac{1}{(2i\Gamma)^3}\frac{-1}{(\xi_{bc}+2i\Gamma)^2}
\right]\nonumber \\
&\quad
-\frac{1}{2}f'''_{-b}\frac{1}{\xi_{bc}(-2i\Gamma)^2}
-f_-''(\xi_b-i\Gamma)\left[
\frac{-1}{\xi_{bc}^2(-2i\Gamma)^2}
+\frac{-2}{\xi_{bc}(-2i\Gamma)^3}
\right] \nonumber \\
&\quad
-\frac{1}{2}f_-'(\xi_b-i\Gamma)\left[
\frac{2}{\xi_{bc}^3(-2i\Gamma)^2}
+\frac{4}{\xi_{bc}^2(-2i\Gamma)^3}
+\frac{6}{\xi_{bc}(-2i\Gamma)^4}
\right]
\Biggr\} \nonumber \\
&+O(\Gamma) \nonumber \\
&=
\frac{3}{16\Gamma^2}\frac{f'_b}{\xi_{bc}}
+\frac{i}{16\Gamma}\frac{f''_b}{\xi_{bc}}
-\frac{i}{4\Gamma}\frac{f'_b}{\xi_{bc}^2}
-\frac{1}{4}\frac{f'_b}{\xi_{bc}^3}
+\frac{1}{32}\frac{f'''_b}{\xi_{bc}}+O(\Gamma),\\
&I_{bbc}^{(2)}\nonumber \\
&= \Gamma^2\Biggl\{\frac{1}{2}f'''_{+b}\frac{1}{(2i\Gamma)^2(\xi_{bc}+2i\Gamma)}\nonumber \\
&\qquad+f_+''(\xi_b+i\Gamma)\left[\frac{-2}{(2i\Gamma)^3(\xi_{bc}+2i\Gamma)}+\frac{-1}{(2i\Gamma)^2(\xi_{bc}+2i\Gamma)^2}\right] \nonumber \\ 
&\quad +\frac{1}{2}f_+'(\xi_b+i\Gamma)\Big[\frac{6}{(2i\Gamma)^4(\xi_{bc}+2i\Gamma)}+\frac{4}{(2i\Gamma)^3(\xi_{bc}+2i\Gamma)^2} \nonumber \\
&\hspace{4cm}+\frac{2}{(2i\Gamma)^2(\xi_{bc}+2i\Gamma)^3}\Big] \nonumber \\
&\quad -f_-''(\xi_b-i\Gamma)\frac{1}{\xi_{bc}(-2i\Gamma)^3}\nonumber \\
&\quad+f_-'(\xi_b-i\Gamma)\left[\frac{1}{\xi_{bc}^2(-2i\Gamma)^3}+\frac{3}{\xi_{bc}(-2i\Gamma)^4}\right] \Biggr\}+O(\Gamma) \nonumber \\
&=
\frac{3}{16\Gamma^2}\frac{f'_b}{\xi_{bc}}
-\frac{i}{16\Gamma}\frac{f''_b}{\xi_{bc}}
-\frac{i}{8\Gamma}\frac{f'_b}{\xi_{bc}^2}
-\frac{1}{8}\frac{f''_b}{\xi_{bc}^2}
+\frac{1}{32}\frac{f'''_b}{\xi_{bc}}+O(\Gamma).
\end{align}

\subsubsection{Case $a\neq b=c$}
Similarly, when $b=c$, the poles at $\xi_b$ merge and higher-order residues must be evaluated.
Expanding these residues up to $O(\Gamma^0)$ gives
\begin{align}
  I_{abb}^{(1)}
=& -\Gamma^2\left[
\frac{f_+'(\xi_b+i\Gamma)}{(\xi_{ba}+2i\Gamma)^2(2i\Gamma)^2\xi_{ba}}
+\frac{f_+'(\xi_a+i\Gamma)}{(2i\Gamma)^2(\xi_{ab}+2i\Gamma)^2\xi_{ab}}\right. \nonumber \\
&\left. 
+\frac{f_-'(\xi_b-i\Gamma)}{\xi_{ba}^2(\xi_{ba}-2i\Gamma)(-2i\Gamma)^2}
+\frac{f_-'(\xi_a-i\Gamma)}{\xi_{ab}^2(-2i\Gamma)^2(\xi_{ab}-2i\Gamma)}
\right] +O(\Gamma) \nonumber \\
= &
f'_b\frac{1}{4\xi_{ba}^3}
+
f'_a\frac{1}{4\xi_{ab}^3}+O(\Gamma),
\end{align}
\begin{align}
  &I_{abb}^{(2)}\nonumber \\
= &\Gamma^2\Bigg\{
f_+''(\xi_b+i\Gamma)\frac{1}{(\xi_{ba}+2i\Gamma)(2i\Gamma)^2\xi_{ba}} \nonumber \\
&\qquad
-f_+'(\xi_b+i\Gamma)\left[
\frac{1}{(\xi_{ba}+2i\Gamma)^2(2i\Gamma)^2\xi_{ba}}\right. \nonumber \\
&\hspace{1cm}\left.
+\frac{2}{(\xi_{ba}+2i\Gamma)(2i\Gamma)^3\xi_{ba}}
+\frac{1}{(\xi_{ba}+2i\Gamma)(2i\Gamma)^2\xi_{ba}^2}
\right]\nonumber \\
&\hspace{2cm}-c.c.\Bigg\}+O(\Gamma)\nonumber \\
=& -\frac{i}{\Gamma}f'_b\frac{1}{4\xi_{ba}^2}+O(\Gamma).
\end{align}

\subsubsection{Case $a=c\neq b$}
When $a=c$, the poles at $\xi_a$ merge and higher-order residues must be evaluated.
Expanding these residues up to $O(\Gamma^0)$ gives
\begin{align}
  &I_{aba}^{(1)}\nonumber \\
=& -\Gamma^2\left\{
\frac{f_+'(\xi_a+i\Gamma)}{(2i\Gamma)^3(\xi_{ab}+2i\Gamma)\xi_{ab}}
+\frac{f''_-(\xi_a-i\Gamma)}{\xi_{ab}(-2i\Gamma)^2(\xi_{ab}-2i\Gamma)} \right. \nonumber \\
&-\frac{1}{2}f_-'(\xi_a-i\Gamma)\left[
2\frac{1}{\xi_{ab}^2(-2i\Gamma)^2(\xi_{ab}-2i\Gamma)}\right. \nonumber \\
&\quad\left.\left.
+2\frac{1}{\xi_{ab}(-2i\Gamma)^2(\xi_{ab}-2i\Gamma)^2}
+2\frac{1}{\xi_{ab}(-2i\Gamma)^3(\xi_{ab}-2i\Gamma)}
\right]
\right\}\nonumber\\
&\hspace{2cm}+O(\Gamma)\nonumber \\
=&-\frac{i}{\Gamma}\frac{f'_a}{8\xi_{ab}^2}+f''_a\frac{1}{8\xi_{ab}^2}-f'_a\frac{1}{4\xi_{ab}^3}+O(\Gamma),
\end{align}

\begin{align}
  &I_{aba}^{(2)}\nonumber \\
=&
\Gamma^2\left\{
-\frac{f_+'(\xi_b+i\Gamma)}{(\xi_{ba}+2i\Gamma)^2(-2i\Gamma)^2\xi_{ba}}
+\frac{f_+'(\xi_a+i\Gamma)}{(2i\Gamma)^2(\xi_{ab}+2i\Gamma)\xi_{ab}^2}\right. \nonumber \\
&\left.
-\frac{f_-'(\xi_b-i\Gamma)}{\xi_{ba}^2(\xi_{ba}-2i\Gamma)(-2i\Gamma)^2}
+\frac{f_-'(\xi_a-i\Gamma)}{\xi_{ab}^2(-2i\Gamma)^2(\xi_{ab}-2i\Gamma)}
\right\} +O(\Gamma) \nonumber \\
=&
f'_b\frac{1}{4\xi_{ba}^3}
-f'_a\frac{1}{4\xi_{ab}^3}+O(\Gamma).
\end{align}

\subsubsection{Case $a=b=c$}
When $a=b=c$, the pole order is maximal.
Keeping terms through $O(\Gamma^0)$, we obtain
\begin{align}
&I_{aaa}^{(1)}\nonumber \\
=&-\Gamma^2 \left\{
  f_+'(\xi_a+i\Gamma)\frac{-4}{(2i\Gamma)^5}+f_+''(\xi_a+i\Gamma)\frac{1}{(2i\Gamma)^4}\right. \nonumber \\
  &-\frac{1}{6}\left[
    \frac{f_-''''(\xi_a-i\Gamma)}{(-2i\Gamma)^2}+\frac{-6f_-'''(\xi_a-i\Gamma)}{(-2i\Gamma)^3}\right. \nonumber \\
  &\quad  \left. \left.  +\frac{18f_-''(\xi_a-i\Gamma)}{(-2i\Gamma)^4}+\frac{-24f_-'(\xi_a-i\Gamma)}{(-2i\Gamma)^5}
  \right]
\right\}\nonumber \\
=&\frac{1}{8i\Gamma^3}f_a' +\frac{1}{16\Gamma^2}f''_a+\frac{1}{96}f''''_a+O(\Gamma)
\end{align}

\begin{align}
&I_{aaa}^{(2)}\nonumber \\
=&\Gamma^2\left[
  \frac{\frac{1}{2}f_+'''(\xi_a+i\Gamma)}{(2i\Gamma)^3}+\frac{-3f_+''(\xi_a+i\Gamma)}{(2i\Gamma)^4}+\frac{6f_+'(\xi_a+i\Gamma)}{(2i\Gamma)^5}
\right]-c.c. \nonumber \\
=&\frac{3}{16i\Gamma^3}f'_a+\frac{1}{32i\Gamma}f'''_a+O(\Gamma)
\end{align}

\begin{align}
  &I_{aa}^{(3)}\nonumber \\
=&-\Gamma^2\left\{\frac{f_+''(\xi_a+i\Gamma)}{(2i\Gamma)^3}+\frac{-3f_+'(\xi_a+i\Gamma)}{(2i\Gamma)^4}\right.\nonumber \\
& -\frac{1}{2}f'''_-(\xi_a-i\Gamma)\frac{1}{(-2i\Gamma)^2}-f''_-(\xi_a-i\Gamma)\frac{-2}{(-2i\Gamma)^3}\nonumber \\
&\left.\hspace{1cm}-f_-'(\xi_a-i\Gamma)\frac{3}{(-2i\Gamma)^4}\right\} \nonumber \\
=&\frac{3}{16\Gamma^2}f'_a+\frac{i}{16\Gamma}f''_a+\frac{1}{32}f'''_a+O(\Gamma).
\end{align}

\subsection{Derivation of each term}
Therefore, collecting the contributions from the five cases, we obtain
\begin{widetext}
\begin{align}
  \sigma^{\mu\alpha\beta}=\frac{e^3}{\hbar}\int \frac{d^d k}{(2\pi)^d}\Bigg\{&\sum_{a\neq b,b\neq c,c\neq a}\left[H_{bc}^\mu H_{ca}^\beta H_{ab}^\alpha \left(f'_a\frac{1}{4\xi_{ab}^2\xi_{ac}}+f'_b\frac{1}{4\xi_{ab}^2\xi_{bc}} \right)+(c.c.\ a\leftrightarrow b)\right] \nonumber\\
  &+\sum_{a\neq b}\left[H_{bb}^\mu H_{ba}^\beta H_{ab}^\alpha \left(-\frac{i}{\Gamma}f'_b\frac{1}{4\xi_{ba}^2}+\frac{f'_b-f'_a}{4\xi_{ba}^3}\right)+(c.c.\ a\leftrightarrow b)\right] \nonumber\\
  &+\sum_{a\neq b}\left[H_{ba}^\mu H_{aa}^\beta H_{ab}^\alpha \left(-\frac{i}{\Gamma}f'_a\frac{1}{8\xi_{ab}^2}+f''_a\frac{1}{8\xi_{ab}^2}-f'_b\frac{1}{4\xi_{ab}^3}-f'_a\frac{1}{2\xi_{ab}^3}\right)+(c.c.\ a\leftrightarrow b)\right] \nonumber\\
  &+\sum_{c\neq b}\left[H_{bc}^\mu H_{cb}^\beta H_{bb}^\alpha \left(\frac{3}{8\Gamma^2}\frac{f'_b}{\xi_{bc}}-\frac{3i}{8\Gamma}\frac{f'_b}{\xi_{bc}^2}-\frac{1}{8}\frac{f''_b}{\xi_{bc}^2}-\frac{1}{4}\frac{f'_b}{\xi_{bc}^3}+\frac{1}{16}\frac{f'''_b}{\xi_{bc}}\right)+c.c.\right]\nonumber\\
  &+\sum_{a\neq b}\left[H_{ba}^{\mu\beta} H_{ab}^\alpha \frac{f'_a}{4\xi_{ab}^2}+(c.c.\ a\leftrightarrow b)\right] \nonumber\\
  &+\sum_a\left[H_{aa}^\mu H_{aa}^\beta H_{aa}^\alpha \left(\frac{1}{16\Gamma^2}f''_a+\frac{1}{96}f''''_a\right)+c.c.\right]\nonumber \\
  &+\sum_a\left[H_{aa}^{\mu\beta} H_{aa}^\alpha \left(\frac{3}{16\Gamma^2}f'_a+\frac{i}{16\Gamma}f''_a+\frac{1}{32}f'''_a\right)+c.c.\right]\Bigg\}+(\alpha \leftrightarrow \beta)+O(\Gamma). \label{eq:nonlinearConductivityUptoTau0}
\end{align}
\end{widetext}

\clearpage

\subsubsection{Drude term}\label{app:NLD}
Collecting the $O(\Gamma^{-2})$ contributions, we obtain
\begin{align}
  &\sigma\smrm{Drude}^{\mu\beta\alpha}\nonumber \\
  =&\frac{1}{\Gamma^2}\frac{e^3}{\hbar}\int \frac{d^d k}{(2\pi)^d}\Biggl[\sum_b\left(
    \frac{3}{8}H_{bb}^{\mu\beta}H_{bb}^\alpha f_b' +\frac{1}{8}H_{bb}^\mu H_{bb}^\beta H_{bb}^\alpha f_b'' \right. \nonumber \\
   &\left. \hspace{2cm} +\sum_{a\neq b}\frac{3}{8}\frac{H_{ba}^\mu H_{ab}^\beta H_{bb}^\alpha +H_{ab}^\mu H_{bb}^\alpha H_{ba}^\beta}{\xi_{ba}}f_b'
  \right) \nonumber \\
  &+(\alpha \leftrightarrow \beta)\Biggr] \nonumber 
\end{align}
\begin{align}
  =&\frac{1}{\Gamma^2}\frac{e^3}{\hbar}\int \frac{d^d k}{(2\pi)^d}\Biggl[\sum_b\left(\frac{3}{8}\partial_{k_\mu} H_{bb}^\beta H_{bb}^\alpha f_b' +\frac{1}{8}H_{bb}^\mu H_{bb}^\beta H_{bb}^\alpha f_b'' \right) \nonumber \\
  &+(\alpha \leftrightarrow \beta) \Biggr]\nonumber \\
  =&-\frac{1}{4\Gamma^2}\frac{e^3}{\hbar}\int \frac{d^d k}{(2\pi)^d}\sum_b\partial_{k_\mu} \partial_{k_\beta} \partial_{k_\alpha} \epsilon_b f_b.
\end{align}
In going to the last line, we used $\partial_{k_\mu} \partial_{k_\alpha} f_b=\partial_{k_\mu} H_{bb}^\alpha f_b' +H_{bb}^\mu H_{bb}^\alpha f_b''$.
This term is called the nonlinear Drude term, which is an $O(\tau^2)$ contribution to the longitudinal nonlinear conductivity.
In Table~\ref{tab:comparison} of the main text, it is referred to as the nonlinear Drude term.

\subsubsection{BCD term}
Collecting the $O(\Gamma^{-1})$ contributions, we obtain
\begin{align}
&\sigma\smrm{BCD}^{\mu\alpha\beta}\nonumber\\
=&\frac{1}{\Gamma}\frac{e^3}{\hbar}\int \frac{d^d k}{(2\pi)^d}\sum_b\Biggl(\frac{3}{8}i\sum_{a\neq b}\frac{-H_{ba}^\mu H_{ab}^\beta H_{bb}^\alpha+H_{ab}^\mu H_{bb}^\alpha H_{ba}^\beta}{\xi_{ba}^2}f_b' \nonumber \\
&\qquad +\frac{1}{8}i\sum_{a\neq b}\frac{-2H_{bb}^\mu H_{ba}^\beta H_{ab}^\alpha+H_{ba}^\mu H_{ab}^\alpha H_{bb}^\beta}{\xi_{ba}^2}f_b' \nonumber \\
&\qquad +\frac{1}{8}i\sum_{a\neq b}\frac{-H_{ba}^\mu H_{aa}^\beta H_{ab}^\alpha+2H_{aa}^\mu H_{ab}^\alpha H_{ba}^\beta}{\xi_{ba}^2}f_a'\Biggl)+(\alpha \leftrightarrow \beta) \nonumber \\
=&\frac{1}{\Gamma}\frac{e^3}{\hbar}\int \frac{d^d k}{(2\pi)^d}\sum_b \frac{1}{8}(3\Omega_b^{\mu\beta}\partial_{k_\alpha} f_b+2\Omega_b^{\beta\alpha}\partial_{k_\mu} f_b+\Omega_b^{\alpha\mu}\partial_{k_\beta} f_b)\nonumber \\
&\hspace{6cm} +(\alpha \leftrightarrow \beta) \nonumber \\
=&\frac{1}{4\Gamma}\frac{e^3}{\hbar}\int \frac{d^d k}{(2\pi)^d}\sum_b (\Omega_b^{\mu\beta}\partial_{k_\alpha} f_b+\Omega_b^{\mu\alpha}\partial_{k_\beta} f_b),
\end{align}
where
\begin{align}
\Omega_b^{\alpha\beta}=&\sum_{a\neq b}\frac{H_{ba}^\alpha H_{ab}^\beta-H_{ba}^\beta H_{ab}^\alpha}{i\xi_{ba}^2}. 
\end{align}
This is called the Berry curvature dipole term, which is an $O(\tau)$ contribution to the nonlinear Hall conductivity but does not contribute to the longitudinal nonlinear conductivity.

\subsubsection{Quantum metric dipole term}\label{app:QMD}
Collecting the $O(\Gamma^0)$ contributions, we obtain
\begin{align}
&\sigma_{\Gamma^0}^{\mu\alpha\beta}\nonumber \\
=&\frac{e^3}{\hbar}\int \frac{d^d k}{(2\pi)^d}\left\{\sum_{b\neq a}\left[\partial_\mu (H_{ba}^\beta H_{ab}^\alpha) \frac{f'_b}{4\xi_{ab}^2}\right.\right.\nonumber \\
&-(H_{bb}^\beta-H_{aa}^\beta)(H_{ba}^\mu H_{ab}^\alpha+H_{ba}^\alpha H_{ab}^\mu) \frac{f'_b}{4\xi_{ab}^3}\nonumber\\
  &+H_{ba}^\mu H_{ab}^{\beta} H_{bb}^\alpha \left(-\frac{f'_b}{4\xi_{ba}^3}\right) +H_{ab}^\mu H_{bb}^{\alpha} H_{ba}^\beta \left(-\frac{f'_b}{4\xi_{ba}^3}\right) \nonumber\\
   &+H_{bb}^\mu H_{ba}^\beta H_{ab}^\alpha \left(\frac{f'_a}{4\xi_{ab}^3}-\frac{f'_b}{4\xi_{ab}^3}\right) +H_{ba}^\mu H_{ab}^{\alpha} H_{bb}^\beta \left(\frac{f'_a}{4\xi_{ab}^3}+\frac{f'_b}{2\xi_{ab}^3}\right)\nonumber\\
  &\left.+H_{ab}^\mu H_{bb}^\beta H_{ba}^\alpha \left(\frac{f'_a}{4\xi_{ab}^3}+\frac{f'_b}{2\xi_{ab}^3}\right)  +H_{bb}^\mu H_{ba}^{\alpha} H_{ab}^\beta \left(-\frac{f'_b}{4\xi_{ab}^3}+\frac{f'_a}{4\xi_{ab}^3}\right)\right]\nonumber\\
  &\left.+\sum_{a}\left[H_{aa}^\mu H_{aa}^\beta H_{aa}^\alpha \frac{f''''_a}{48}+ \partial_\mu H_{aa}^{\beta} H_{aa}^\alpha \frac{f'''_a}{16}\right] \right\}+(\alpha \leftrightarrow \beta) .
\end{align}
Here, we used
\begin{align}
  &\left\{\sum_{a\neq b,b\neq c,c\neq a}\left[H_{bc}^\mu H_{ca}^\beta H_{ab}^\alpha \left(f'_a\frac{1}{4\xi_{ab}^2\xi_{ac}}+f'_b\frac{1}{4\xi_{ab}^2\xi_{bc}} \right)\right.\right. \nonumber \\
  &\left. \hspace{5cm}+(c.c.\ a\leftrightarrow b)\right]\nonumber \\
  &\left.+\sum_{a\neq b}\left[H_{ba}^{\mu\beta} H_{ab}^\alpha \frac{f'_a}{4\xi_{ab}^2}+(c.c.\ a\leftrightarrow b)\right]+(\alpha\leftrightarrow \beta)\right\}, \nonumber \\
  =&\sum_{a\neq b}\left[\partial_\mu (H_{ba}^\beta H_{ab}^\alpha) \frac{f'_b}{4\xi_{ab}^2}-(H_{bb}^\beta-H_{aa}^\beta)(H_{ba}^\mu H_{ab}^\alpha+c.c.) \frac{f'_b}{4\xi_{ab}^3}\right]\nonumber \\
  &\hspace{6cm}+(\alpha\leftrightarrow \beta),\\
&\sum_{c\neq b}\left(H_{bc}^\mu H_{cb}^\beta H_{bb}^\alpha \frac{1}{16}\frac{f'''_b}{\xi_{bc}}+c.c.\right)+\sum_a \left(H_{aa}^{\mu\beta} H_{aa}^\alpha \frac{1}{32}f'''_a+c.c.\right)\nonumber \\
=&\sum_{a} \partial_\mu H_{aa}^{\beta} H_{aa}^\alpha \frac{f'''_a}{16}, 
\end{align}
and some cancellations between the terms explicitly written and the $(\alpha \leftrightarrow \beta)$ terms among the $O(\Gamma^0)$ contributions in Eq.~\eqref{eq:nonlinearConductivityUptoTau0}.
By defining $g_{ab}^{\alpha\beta}\equiv \frac{1}{2}(H_{ab}^\alpha H_{ba}^\beta +H_{ab}^\beta H_{ba}^\alpha)/{\xi_{ab}^2}$, we obtain
\begin{align}
  &\sigma_{\Gamma^0}^{\mu\alpha\beta}\nonumber \\
=&\frac{e^3}{\hbar}\int \frac{d^d k}{(2\pi)^d}\left\{\sum_{b\neq a}\left[\partial_\mu (H_{ba}^\beta H_{ab}^\alpha+H_{ba}^\alpha H_{ab}^\beta) \frac{f'_b}{4\xi_{ab}^2}\right.\right.\nonumber\\
  &\left.+\left[(H_{bb}^\mu-H_{aa}^\mu )g_{ba}^{\alpha\beta}-2H_{bb}^\mu g_{ba}^{\alpha\beta}+H_{bb}^\beta g_{ba}^{\mu\alpha}+H_{bb}^\alpha g_{ba}^{\mu\beta} \right]\frac{f'_b}{\xi_{ab}}\right] \nonumber\\
  &\left.+\sum_{a}\left[H_{aa}^\mu H_{aa}^\beta H_{aa}^\alpha \frac{f''''_a}{24}+ (\partial_\mu H_{aa}^{\beta} H_{aa}^\alpha +\partial_\mu H_{aa}^{\alpha} H_{aa}^\beta )\frac{f'''_a}{16}\right]\right\}.
\end{align}
Finally, using the relation
\begin{align}
    \partial_\mu g_{ba}^{\beta\alpha}=&\partial_\mu (H_{ab}^\beta H_{ba}^\alpha +H_{ab}^\alpha H_{ba}^\beta)\frac{1}{2\xi_{ab}^2}\nonumber \\
    &+(H_{bb}^\mu-H_{aa}^\mu)(H_{ab}^\beta H_{ba}^\alpha +H_{ab}^\alpha H_{ba}^\beta)\frac{1}{\xi_{ab}^3}\\
  \partial_\mu [H_{aa}^\beta H_{aa}^\alpha \frac{f'''_a}{16}]=&(\partial_\mu H_{aa}^\beta H_{aa}^\alpha +H_{aa}^\beta \partial_\mu H_{aa}^\alpha)\frac{f'''_a}{16}\nonumber \\
    &+H_{aa}^\beta H_{aa}^\alpha H_{aa}^\mu \frac{f''''_a}{16}
\end{align}
and periodicity in $k$-space, we obtain
\begin{align}
  \sigma_{\Gamma^0}^{\mu\alpha\beta}
=&\frac{e^3}{\hbar}\int \frac{d^d k}{(2\pi)^d}\left\{\sum_b \partial_\mu g_{b}^{\beta\alpha} \frac{f'_b}{2}\right.\nonumber\\
  &-\sum_{b\neq a}\left[2H_{bb}^\mu g_{ba}^{\alpha\beta}- H_{bb}^\beta g_{ba}^{\mu\alpha}- H_{bb}^\alpha g_{ba}^{\mu\beta} \right]\frac{f'_b}{\xi_{ab}} \nonumber\\
  &\left.-\sum_{a}H_{aa}^\mu H_{aa}^\beta H_{aa}^\alpha \frac{f''''_a}{48}\right\},
\end{align}
where $g_b^{\alpha\beta}=\sum_{a\neq b}g_{ba}^{\alpha\beta}$ is the quantum metric of the $b$-th band.
The first term is the intraband quantum metric dipole term, the second term is the interband quantum metric dipole term, and the last term is the kinetic term~\cite{Ulrich2026,Guo2026}.
The first and last terms contribute to the longitudinal nonlinear conductivity, whereas the second term does not. 
The first term is referred to as the quantum metric dipole term in Table~\ref{tab:comparison} of the main text, and the last term is interpreted as a part of the nonlinear Drude term because $\int \frac{d^d k}{(2\pi)^d}\sum_{a}H_{aa}^\mu H_{aa}^\beta H_{aa}^\alpha f''''_a=\int \frac{d^d k}{(2\pi)^d}\sum_{a}2\partial_\mu \partial_\beta \partial_\alpha \epsilon_a f_a''$.

\section{Derivation of Eq.~\eqref{eq:sigma2}} \label{app:derivationSigma2}
In this section, we derive Eq.~\eqref{eq:sigma2} from Eq.~\eqref{eq:generalNonReciprocalKernel} by setting both $\mu$ and $\beta$ to $\alpha$.
For later convenience, we introduce
\begin{align}
  &\Sigma^{\alpha\alpha\alpha} =-\frac{\Gamma^2 }{2\pi i} \int dx f'(x)\nonumber \\
  &\times \left[\sum_{abc}H_{bc}^\alpha H_{ca}^\alpha H_{ab}^\alpha (G^R_c+G^A_c) \left(G^R_a-G^A_b\right)G^R_a G^A_a G^R_b G^A_b  \right. \nonumber \\
  &\left. +\sum_{ab}H_{ba}^{\alpha\alpha} H_{ab}^\alpha\left(G^R_a-G^A_b\right)G^R_a G^A_a G^R_b G^A_b \right]. \label{eq:longitudinalNonReciprocalKernel}
\end{align}
as the momentum-space contribution of $\sigma^{\alpha\alpha\alpha}$, i.e., $\sigma^{\alpha\alpha\alpha}=\frac{2e^3}{\hbar}\int \frac{d^d k}{(2\pi)^d}\Sigma^{\alpha\alpha\alpha}$.
From now on, we assume TR symmetry $[T,H(k)]=0, \ T=KU(\bm{k}\to -\bm{k})$ with $K$ being the complex conjugation operator and $U$ being a unitary matrix, which implies 
\begin{align}
  \mel{a}{\partial_{k_\alpha} H }{b}
  &=-\mel{b}{\partial_{k_\alpha} H }{a}|_{\bm{k}\to -\bm{k}} \\
  \mel{a}{\partial_{k_\alpha}^2 H }{b}&=\mel{b}{\partial_{k_\alpha}^2 H }{a}|_{\bm{k}\to -\bm{k}}
\end{align}
Because of these relations, both coefficients of $H_{bc}^\alpha H_{ca}^\alpha H_{ab}^\alpha$ and $H_{ba}^{\alpha\alpha} H_{ab}^\alpha$ in Eq.~\eqref{eq:longitudinalNonReciprocalKernel} can be antisymmetrized under the interchange of arbitrary band indices.
Therefore, the coefficient multiplying $H_{bc}^\alpha H_{ca}^\alpha H_{ab}^\alpha$ in Eq.~\eqref{eq:longitudinalNonReciprocalKernel} can be rewritten as
\begin{align}
  &(G^R_c+G^A_c) (G^R_a-G^A_b)G^R_a G^A_a G^R_b G^A_b \nonumber \\
\overset{\mathcal{A}}{=} &(G^R_c+G^A_c) (G^R_a+G^A_a)G^R_a G^A_a G^R_b G^A_b \nonumber \\
=&-\frac{1}{4\Gamma^2}(G^R_c+G^A_c) (G^{R2}_a-G^{A2}_a)(G^{R}_b-G^{A}_b) \nonumber \\
\overset{\mathcal{A}}{=} &-\frac{1}{4\Gamma^2}(-G^R_c G^A_b+G^A_c G^R_b) (G^{R2}_a-G^{A2}_a)\nonumber \\
\overset{\mathcal{A}}{=} &-\frac{1}{2\Gamma^2}(G^R_c G^A_b G^{A2}_a+G^A_c G^R_b G^{R2}_a) \nonumber \\
=&-\frac{1}{2\Gamma^2}(G^R_c G^A_b G^{A2}_a+c.c.),
\end{align}
where $\overset{\mathcal{A}}{=}$ indicates that the expression is equivalent under the antisymmetrization mentioned above.
Therefore, the coefficient multiplying $H_{bc}^\alpha H_{ca}^\alpha H_{ab}^\alpha$ in Eq.~\eqref{eq:longitudinalNonReciprocalKernel} is
\begin{align}
  &\frac{1}{2}\frac{1}{2\pi i}\int dx f'(x)G^R_c G^A_b G^{A2}_a -c.c.\nonumber \\
  =&\frac{1}{2}\left[\frac{f_+'(\xi_b+i\Gamma)}{(\xi_{bc}+2i\Gamma)\xi_{ba}^2}+\frac{f_+''(\xi_a+i\Gamma)}{(\xi_{ac}+2i\Gamma)\xi_{ab}}\right. \nonumber \\
  &-f_+'(\xi_a+i\Gamma)\left(\frac{1}{(\xi_{ac}+2i\Gamma)\xi_{ab}^2}+\frac{1}{(\xi_{ac}+2i\Gamma)^2\xi_{ab}}\right) \nonumber \\
  &\left.-\frac{f_-'(\xi_c-i\Gamma)}{(\xi_{cb}-2i\Gamma)(\xi_{ca}-2i\Gamma)^2}\right]-c.c. \nonumber \\
  \overset{\mathcal{A}}{=}&\frac{1}{2}\left[
    -f_+'(\xi_a+i\Gamma)\left(
      \frac{1}{(\xi_{ac}+2i\Gamma)^2\xi_{ab}}+\frac{2}{(\xi_{ac}+2i\Gamma)\xi_{ab}^2} \right. \right. \nonumber \\
      &\left. \qquad \qquad+\frac{1}{(\xi_{ab}+2i\Gamma)(\xi_{ac}+2i\Gamma)^2}
    \right) \nonumber \\
   & \left. \qquad +f_+''(\xi_a+i\Gamma)\frac{-2i\Gamma}{\xi_{ac} \xi_{ab} (\xi_{ac}+2i\Gamma)} 
  \right]-c.c.
\end{align}
for $a\neq b,\ b\neq c,\ c\neq a$.
In going to the last line, we used $\frac{1}{(\xi_{ac}+2i\Gamma)\xi_{ab}}+\frac{2i\Gamma}{\xi_{ac}\xi_{ab}(\xi_{ac}+2i\Gamma)}=\frac{1}{\xi_{ac}\xi_{ab}}$.

Similarly, the coefficient multiplying $H_{ba}^{\alpha\alpha} H_{ab}^\alpha$ in Eq.~\eqref{eq:longitudinalNonReciprocalKernel} can be rewritten as
\begin{align}
  &(G^R_a-G^A_b)G^R_a G^A_a G^R_b G^A_b \nonumber \\
\overset{\mathcal{A}}{=}&(G^R_a+G^A_a)G^R_a G^A_a G^R_b G^A_b \nonumber \\
=&-\frac{1}{4\Gamma^2}(G^{R2}_a-G^{A2}_a)(G^{R}_b-G^{A}_b) \nonumber \\
=&-\frac{1}{4\Gamma^2}(G^{R2}_a-G^{A2}_a)G^R_b +c.c.
\end{align}
Therefore, the coefficient multiplying $H_{ba}^{\alpha\alpha} H_{ab}^\alpha$ in Eq.~\eqref{eq:longitudinalNonReciprocalKernel} is
\begin{align}
  &\frac{1}{4}\frac{1}{2\pi i}\int dx f'(x)(G^{R2}_a-G^{A2}_a)G^R_b -c.c. \nonumber \\
  =&\frac{1}{4}\left(-\frac{f_+''(\xi_a+i\Gamma)}{\xi_{ab}+2i\Gamma}+\frac{f_+'(\xi_a+i\Gamma)}{(\xi_{ab}+2i\Gamma)^2}+\frac{f_-'(\xi_b-i\Gamma)}{(\xi_{ba}-2i\Gamma)^2} \right. \nonumber \\
      &\left. -\frac{f_-'(\xi_b-i\Gamma)}{\xi_{ba}^2}-\frac{f_-''(\xi_a-i\Gamma)}{\xi_{ab}}+\frac{f_-'(\xi_{a}-i\Gamma)}{\xi_{ab}^2}\right)-c.c. \nonumber \\
\overset{\mathcal{A}}{=}&\frac{1}{2}\left[
  f_+''(\xi_a+i\Gamma)\frac{i\Gamma}{\xi_{ab}(\xi_{ab}+2i\Gamma)} \right. \nonumber \\
&\qquad \left. +f_+'(\xi_a+i\Gamma)\left(\frac{1}{(\xi_{ab}+2i\Gamma)^2}-\frac{1}{\xi_{ab}^2}\right)
\right]-c.c.
\end{align}
for $a\neq b$.

\begin{widetext}
Using the TR symmetry, for arbitrary function $F(a,b)$ with $a,b$ being band indices, we have
\begin{align}
  &\sum_{a\neq b}H_{ba}^{\alpha\alpha} H_{ab}^\alpha F(a,b) \nonumber \\
  =&\sum_{a\neq b}\left\{[\partial_{k_\alpha} \ln H_{ba}^\alpha+(\mel{u_b}{\partial_{k_\alpha}}{u_b}-\mel{u_a}{\partial_{k_\alpha}}{u_a})]H_{ba}^\alpha H_{ab}^\alpha F(a,b)-\sum_{c\neq a, c\neq b} \left(\frac{H_{bc}^\alpha H_{ca}^\alpha}{\xi_{bc}}-\frac{H_{bc}^\alpha H_{ca}^\alpha}{\xi_{ca}}\right)H_{ab}^\alpha F(a,b)\right\}\nonumber \\
  \overset{\mathcal{A}}{=}&\frac{1}{2}\sum_{a\neq b}\left[\partial_{k_\alpha} \ln \frac{H_{ba}^\alpha}{H_{ab}^\alpha}+2(\mel{u_b}{\partial_{k_\alpha}}{u_b}-\mel{u_a}{\partial_{k_\alpha}}{u_a})\right]H_{ba}^\alpha H_{ab}^\alpha F(a,b)-\sum_{a\neq b, b\neq c, c\neq a} H_{bc}^\alpha H_{ca}^\alpha H_{ab}^\alpha\left(\frac{1}{\xi_{bc}}-\frac{1}{\xi_{ca}}\right) F(a,b) \nonumber \\
  = &-i\sum_{a\neq b} R_{ba}H_{ba}^\alpha H_{ab}^\alpha F(a,b)-\sum_{a\neq b, b\neq c, c\neq a} H_{bc}^\alpha H_{ca}^\alpha H_{ab}^\alpha\left(\frac{1}{\xi_{bc}}-\frac{1}{\xi_{ca}}\right) F(a,b)
\end{align}
Therefore, using
\begin{align}
      &-\frac{1}{(\xi_{ac}+2i\Gamma)^2\xi_{ab}}-\frac{2}{(\xi_{ac}+2i\Gamma)\xi_{ab}^2} -\frac{1}{(\xi_{ab}+2i\Gamma)(\xi_{ac}+2i\Gamma)^2}\nonumber \\
      =&-\frac{1}{(\xi_{ac}+2i\Gamma)^2}\left(
        \frac{1}{\xi_{ab}}+\frac{1}{\xi_{cb}}
      \right)-\frac{1}{(\xi_{ac}+2i\Gamma)}\left(\frac{2}{\xi_{ab}^2}-\frac{2}{\xi_{cb}^2}\right) +\frac{1}{(\xi_{ac}+2i\Gamma)^2}\frac{1}{\xi_{cb}}
      -\frac{1}{(\xi_{ac}+2i\Gamma)}\frac{2}{\xi_{cb}^2}\nonumber \\
      &  +\frac{1}{(\xi_{ac}+2i\Gamma)^2\xi_{bc}}-\frac{1}{\xi_{bc}^2}\left(\frac{1}{\xi_{ab}+2i\Gamma}-\frac{1}{\xi_{ac}+2i\Gamma}\right)\nonumber \\
      =&-\frac{1}{\xi_{ac}+2i\Gamma}\left(
        \frac{1}{\xi_{ab}}+\frac{1}{\xi_{cb}}
      \right)\left(\frac{1}{\xi_{ac}+2i\Gamma}+\frac{2}{\xi_{ab}}-\frac{2}{\xi_{cb}}\right)-\frac{1}{\xi_{bc}^2}\left(\frac{1}{\xi_{ab}+2i\Gamma}+\frac{1}{\xi_{ac}+2i\Gamma}\right),
\end{align}
Eq.~\eqref{eq:longitudinalNonReciprocalKernel} can be rewritten as
\begin{align}
  \Sigma^{\alpha\alpha\alpha} =&\sum_{a\neq b}R_{ba}H_{ba}^\alpha H_{ab}^\alpha \Im \left[
  f_+''(\xi_a+i\Gamma)\frac{i\Gamma}{\xi_{ab}(\xi_{ab}+2i\Gamma)}  +f_+'(\xi_a+i\Gamma)\left(\frac{1}{(\xi_{ab}+2i\Gamma)^2}-\frac{1}{\xi_{ab}^2}\right)
\right]\nonumber \\
 &-\sum_{a\neq b, b\neq c, c\neq a} iH_{bc}^\alpha H_{ca}^\alpha H_{ab}^\alpha\left(\frac{1}{\xi_{ac}}+\frac{1}{\xi_{bc}}\right)\Im \left[
  f_+''(\xi_a+i\Gamma)\frac{i\Gamma}{\xi_{ab}(\xi_{ab}+2i\Gamma)}  +f_+'(\xi_a+i\Gamma)\left(\frac{1}{(\xi_{ab}+2i\Gamma)^2}-\frac{1}{\xi_{ab}^2}\right)
\right]\nonumber \\
 &+\sum_{a\neq b, b\neq c, c\neq a} iH_{bc}^\alpha H_{ca}^\alpha H_{ab}^\alpha \Im \left[
  f_+''(\xi_a+i\Gamma)\frac{-2i\Gamma}{\xi_{ac} \xi_{ab} (\xi_{ac}+2i\Gamma)}  \right. \nonumber \\
&\qquad \left. \hspace{5cm}-f_+'(\xi_a+i\Gamma)\frac{1}{\xi_{ac}+2i\Gamma}\left(
        \frac{1}{\xi_{ab}}+\frac{1}{\xi_{cb}}
      \right)\left(\frac{1}{\xi_{ac}+2i\Gamma}+\frac{2}{\xi_{ab}}-\frac{2}{\xi_{cb}}\right)
\right]\nonumber \\
=&\sum_{a\neq b}R_{ba}H_{ba}^\alpha H_{ab}^\alpha \Im \left[
  f_+''(\xi_a+i\Gamma)\frac{i\Gamma}{\xi_{ab}(\xi_{ab}+2i\Gamma)}+f_+'(\xi_a+i\Gamma)\frac{-4i\Gamma\xi_{ab}+4\Gamma^2}{(\xi_{ab}+2i\Gamma)^2\xi_{ab}^2} 
\right]\nonumber \\
 &+\sum_{a\neq b, b\neq c, c\neq a} iH_{bc}^\alpha H_{ca}^\alpha H_{ab}^\alpha\Im \left[
  f_+''(\xi_a+i\Gamma)\frac{-i\Gamma}{\xi_{ac}\xi_{bc}(\xi_{ab}+2i\Gamma)}+f_+'(\xi_a+i\Gamma)\left(\frac{1}{\xi_{ac}}+\frac{1}{\xi_{bc}}\right)\frac{4i\Gamma}{\xi_{ac}\xi_{bc}(\xi_{ab}+2i\Gamma)}
\right]
\end{align}
where we used $\left(\frac{1}{\xi_{ac}}+\frac{1}{\xi_{bc}}\right)\left(\frac{1}{\xi_{ab}^2}+\left(\frac{2}{\xi_{ac}}-\frac{2}{\xi_{bc}}\right)\frac{1}{\xi_{ab}}\right)-(b \leftrightarrow c)=0$ for the last equality.
Using $H_{ab}^\alpha=\xi_{ab}i A_{ab}^\alpha$ and Eq.~\eqref{eq:sigmaMuAlphaBetaGreenFunction}, we obtain Eq.~\eqref{eq:sigma2}.

\end{widetext}

\section{Derivation of Eqs.~\eqref{eq:skewness} and \eqref{eq:skewness-low-temp}}\label{app:skewnessLowTemp}
In this section, we derive Eqs.~\eqref{eq:skewness} and \eqref{eq:skewness-low-temp} from Eq.~\eqref{eq:sigma2}, assuming that the bands are well separated $|\Delta_{ab}|> \Delta\smrm{min} \gg \Gamma$ for all $a\neq b$.

First, we consider the case $\beta \Gamma \ll 2\pi$, which enables us to expand $D^{(2)}_{ab}$ and $D^{(3)}_{abc}$ in $\Gamma$ and obtain $D^{(2)}_{ab}=\Gamma \left(\frac{4}{\Delta_{ab}}f_a'-f_a''\right)+O(\Gamma^2),\ D^{(3)}_{abc}=\Gamma \left[-4\left(\frac{1}{\Delta_{ac}}+\frac{1}{\Delta_{bc}}\right)f_a'+f_a''\right]+O(\Gamma^2)$.
In this situation, we further consider the low-temperature regime in which the chemical potential is offset from the band edge by an energy of order $\beta^{-1}$, i.e. $\Delta\smrm{min}\gg |\xi_a|\sim \beta^{-1}(\gg \Gamma)$.
Assuming this situation, we obtain
\begin{align}
  D^{(2)}_{ab}\simeq -\Gamma f''_{a},\
  D^{(3)}_{abc}\simeq \Gamma f''_{a}.
\end{align}
By substituting these expressions into Eq.~\eqref{eq:sigma2}, we obtain
\begin{align}
  \sigma^{xxx}\simeq&\frac{e^3}{\hbar}\int_{\mathrm{BZ}}\frac{d^d k}{(2\pi)^d}\sum_{a}^{\xi_a\sim \beta^{-1}}\Gamma f''_{a}\nonumber \\
  &\qquad  \times\Bigg[-\sum_{b\ne a}R_{ab}|A_{ab}|^2
  +	\sum_{\substack{b\ne a\\c\ne a,\ c\ne b}}\Re[A_{ab}A_{bc}A_{ca}]\Bigg]
\end{align}
The term in the square brackets can be rewritten as
\begin{align}
  &-\sum_{b\ne a}R_{ab}|A_{ab}|^2+	\sum_{\substack{b\ne a\\c\ne a,\ c\ne b}}\Re[A_{ab}A_{bc}A_{ca}]\\
  =&\Re \sum_{b\ne a}A_{ab}\left[A_{ba}(A_{bb}-A_{aa})+i\partial_{k_\alpha} A_{ba} +\sum_{c\neq a,b}A_{bc}A_{ca}\right]\nonumber \\
  =&\Re \sum_{b\ne a}A_{ab}\left[-2A_{ba}A_{aa}+i\partial_{k_\alpha} A_{ba} +\sum_{c}A_{bc}A_{ca}\right]\nonumber \\
  =&\Re \sum_{b\ne a}i\braket{u_a}{\partial_{k_\alpha} u_b}\left[
    2\braket{u_b}{\partial_{k_\alpha}u_a}\braket{u_a}{\partial_{k_\alpha}u_a}-\braket{u_b}{\partial_{k_\alpha}^2u_a}
  \right]\nonumber \\
  =&\Re \left[(-i\partial_{k_\alpha}-A_{aa})\bra{u_a}\right]\left[(i\partial_{k_\alpha}-A_{aa})^2\ket{u_a}\right]\nonumber \\
  =&\Re \mel{u_a}{(i\partial_{k_\alpha}-A_{aa})^3}{u_a} \label{eq:skewnessRelation}
\end{align}
Therefore, we obtain Eq.~\eqref{eq:skewness}.
This is the geometric skewness of the wave packet $\sim \mel{u_a}{(\Delta r_\alpha)^3}{u_a}$, which indicates the asymmetry of the real-space distribution of the wave packet.
When system has inversion symmetry $\ket{u_a(\bm k)}=e^{i\theta_a(-\bm k)}U\ket{u_a(-\bm k)}$ where $U$ is an arbitrary $k$-independent unitary matrix and $\theta_a(\bm k)$ is a gauge-dependent phase, using the relation $A_{aa}(-\bm{k})=-A_{aa}(\bm k)+\partial_{k_\alpha}\theta_a(\bm k)$, we obtain 
\begin{align}
  &\Re \mel{u_a(-\bm k)}{(i\partial_{-k_\alpha}-A_{aa}(-\bm k))^3}{u_a(-\bm k)}\nonumber \\
  =&-\Re \mel{u_a(\bm k)}{(i\partial_{k_\alpha}-A_{aa}(\bm k))^3}{u_a(\bm k)}.
\end{align}
Thus, the skewness becomes $k$-odd function when the system has inversion symmetry.
On the other hand, when the system has TR symmetry $\ket{u_a(\bm k)}=e^{i\theta_a(-\bm k)}U\ket{u_a^*(-\bm k)}$, using the relation $A_{aa}(-\bm{k})=A_{aa}(\bm k)+\partial_{k_\alpha}\theta_a(\bm k)$, we obtain
\begin{align}
  &\Re \mel{u_a(-\bm k)}{(i\partial_{-k_\alpha}-A_{aa}(-\bm k))^3}{u_a(-\bm k)}\nonumber \\
  =&\Re \mel{u_a(\bm k)}{(i\partial_{k_\alpha}-A_{aa}(\bm k))^3}{u_a(\bm k)}.
\end{align}
Thus, the skewness becomes $k$-even function when the system has TR symmetry.
Therefore, in the TR-symmetric systems considered here, a nonzero value of this quantity provides a geometric signature of inversion-symmetry breaking.

Next, we consider the case $\beta \Gamma \gg 2\pi$, which enables us to use the asymptotic form of the digamma function $\psi(z) \sim \log z$.
In this case, we obtain
\begin{align}
  D^{(2)}_{ab}=&\frac{-4\Gamma^2(\Delta_{ab}^3+\xi_a(3\Delta_{ab}^2+4\Gamma^2))}{\pi(\Delta_{ab}^2+4\Gamma^2)^2(\xi_a^2+\Gamma^2)}\nonumber \\
  &\qquad    -\frac{2\Gamma\Delta_{ab}(\Gamma \xi_a\Delta_{ab}+\Gamma(\xi_a^2-\Gamma^2))}{\pi(\xi_a^2+\Gamma^2)^2(\Delta_{ab}^2+4\Gamma^2)}\\
  D^{(3)}_{abc}=&\left(\frac{1}{\Delta_{ac}}+\frac{1}{\Delta_{bc}}\right)\frac{4\Gamma^2\Delta_{ab}(\Delta_{ab}+2\xi_a)}{\pi(\xi_a^2+\Gamma^2)(\Delta_{ab}^2+4\Gamma^2)}\nonumber \\
  &\qquad    +\frac{2\Gamma^2\Delta_{ab}(\Delta_{ab}\xi_a+\xi_a^2-\Gamma^2)}{\pi(\xi_a^2+\Gamma^2)^2(\Delta_{ab}^2+4\Gamma^2)}
\end{align}
Using the assumption that the bands are well separated $|\Delta_{ab}|>\Delta\smrm{min}\gg \Gamma$, we obtain
\begin{align}
  D_{ab}^{(2)}=&-\frac{2\Gamma^2 \xi_a}{\pi(\xi_a^2+\Gamma^2)^2}+O\left(\frac{\Gamma}{\Delta\smrm{min}}\right),\\
  D_{abc}^{(3)}=&\frac{2\Gamma^2 \xi_a}{\pi(\xi_a^2+\Gamma^2)^2}+O\left(\frac{\Gamma}{\Delta\smrm{min}}\right)
\end{align}
By substituting these expressions into Eq.~\eqref{eq:sigma2}, we obtain
\begin{align}
  \sigma^{xxx}\simeq&\frac{e^3}{\hbar}\int_{\mathrm{BZ}}\frac{d^d k}{(2\pi)^d}\sum_{a}^{\xi_a\sim \Gamma}\frac{2\Gamma^2 \xi_a}{\pi(\xi_a^2+\Gamma^2)^2}\nonumber \\
  &\qquad  \times\Bigg[-\sum_{b\ne a}R_{ab}|A_{ab}|^2
  +	\sum_{\substack{b\ne a\\c\ne a,\ c\ne b}}\Re[A_{ab}A_{bc}A_{ca}]\Bigg]
\end{align}
By using Eq.~\eqref{eq:skewnessRelation}, we obtain Eq.~\eqref{eq:skewness-low-temp}.

Finally, we show the relation between the skewness and the quantum geometric connection.
Defining the covariant derivative $D_a\equiv \partial_{k_\alpha}+iA_{aa}$, we have
\begin{align}
  \partial_{k_\alpha} P_a \ket{u_a}=&\ket{\partial_{k_\alpha} u_a}+iA_{aa}\ket{u_a}=D_a \ket{u_a}\\
  \partial_{k_\alpha}^2 P_a\ket{u_a}=&(\ketbra{\partial_{k_\alpha}^2 u_a}{u_a}+2\ketbra{\partial_{k_\alpha} u_a}{\partial_{k_\alpha} u_a}+\ketbra{u_a}{\partial_{k_\alpha}^2 u_a})\ket{u_a}\nonumber \\
  =&\ket{\partial_{k_\alpha}^2 u_a}+2iA_{aa}\ket{\partial_{k_\alpha} u_a}+\braket{\partial_{k_\alpha}^2 u_a}{u_a}\ket{u_a}\\
  D_a^2\ket{u_a}=&(\partial_{k_\alpha}+iA_{aa})^2\ket{u_a}\nonumber\\
  =&\ket{\partial_{k_\alpha}^2 u_a}+i\partial_{k_\alpha} A_{aa}\ket{u_a}+2iA_{aa}\ket{\partial_{k_\alpha} u_a}-A_{aa}^2\ket{u_a}\nonumber\\
  =&\partial_{k_\alpha}^2 P_a\ket{u_a}+(i\partial_{k_\alpha} A_{aa}-A_{aa}^2-\braket{\partial_{k_\alpha}^2 u_a}{u_a})\ket{u_a}\nonumber\\
  =&\partial_{k_\alpha}^2 P_a\ket{u_a}+g_a\ket{u_a}
\end{align}
where $g_a\equiv \sum_{b\neq a}A_{ab}A_{ba}$ is the quantum metric of the $a$-th band.
Since $\braket{u_a}{D_a u_a}=0$, we obtain
\begin{align}
  C_{\alpha\alpha\alpha}^a\equiv&\Tr [P_a \partial_{k_\alpha} P_a \partial_{k_\alpha}^2 P_a]\nonumber\\
  =&\bra{u_a}\partial_{k_\alpha} P_a \partial_{k_\alpha}^2 P_a\ket{u_a}\nonumber\\
  =&\bra{D_a u_a}\ket{D_a^2 u_a}\nonumber\\
  =&-\partial_{k_\alpha} g_a-i\mel{u_a}{(i\partial_{k_\alpha}-A_{aa})^3}{u_a}
\end{align}
Taking the imaginary part of both sides, we obtain the direct relationship between the quantum geometric connection and the skewness of the Bloch wave function, 
\begin{align}
  -\Im C_{\alpha\alpha\alpha}^a=\Re \mel{u_a}{(i\partial_{k_\alpha}-A_{aa})^3}{u_a}.
\end{align}

\section{Leading order contribution in $\Gamma$}\label{app:leadingOrderGamma}
In this section, we identify the leading order contribution of the nonreciprocal current in $\Gamma$.
We first consider the case where the chemical potential lies within the band gap, which applies to insulators and semiconductors with a finite density of charge carriers.
In this case, there exists a minimum nonzero value $\xi\smrm{min}$ such that $\xi\smrm{min}\leq |\xi_a|$ for all $a$ where $\xi_a\equiv \epsilon_a-\mu$, and thus we can consider the low temperature regime $\Gamma\ll \xi\smrm{min},\ \beta \Gamma \gg 2\pi$, where we can use the asymptotic form of the digamma function $\psi(z) \sim \log z$.
In this regime, we obtain 
\begin{align}
  D^{(2)}_{ab}=&2\Gamma^2 \frac{-10\xi_a^2+5\xi_a\xi_b-\xi_b^2}{\pi \Delta_{ab}^2 \xi_a^3}+O(\Gamma^3), \label{eq:lowTmpGammaSquared}\\
  D^{(3)}_{abc}=&\frac{2\Gamma^2}{\pi\Delta_{ab}\xi_a^2} \Bigg[\left(\frac{1}{\Delta_{ac}}+\frac{1}{\Delta_{bc}}\right)(2\Delta_{ab}+4\xi_a)+\frac{\Delta_{ab}}{\xi_a}+1\Bigg]+O(\Gamma^3).
\end{align}
Thus the leading order contribution to $\sigma^{\alpha\alpha\alpha}$ is $O(\Gamma^2)$.
In the high temperature regime $\beta \Gamma \ll 2\pi$, we obtain
\begin{align}
  D^{(2)}_{ab}=&\Gamma \left(\frac{4}{\Delta_{ab}}f_a'-f_a''\right)+O(\Gamma^2), \label{eq:highTmpGammaLinear}\\
  D^{(3)}_{abc}=&\Gamma \left[-\left(\frac{1}{\Delta_{ac}}+\frac{1}{\Delta_{bc}}\right)4f_a'+f_a''\right]+O(\Gamma^2).
\end{align}
Thus the leading order contribution to $\sigma^{\alpha\alpha\alpha}$ is $O(\Gamma)$.
On the other hand, when the chemical potential $\mu$ lies within the conduction band (ordinary metallic cases), there is a contribution from the states satisfying $\epsilon_a-\mu=0$.
Therefore, we cannot consider the low temperature regime $\beta \Gamma \gg 2\pi$ based on a finite lower bound of $|\xi_a|$, and the leading order contribution to $\sigma^{\alpha\alpha\alpha}$ is $O(\Gamma)$ in this case.

\begin{figure}
  \includegraphics[width=\linewidth]{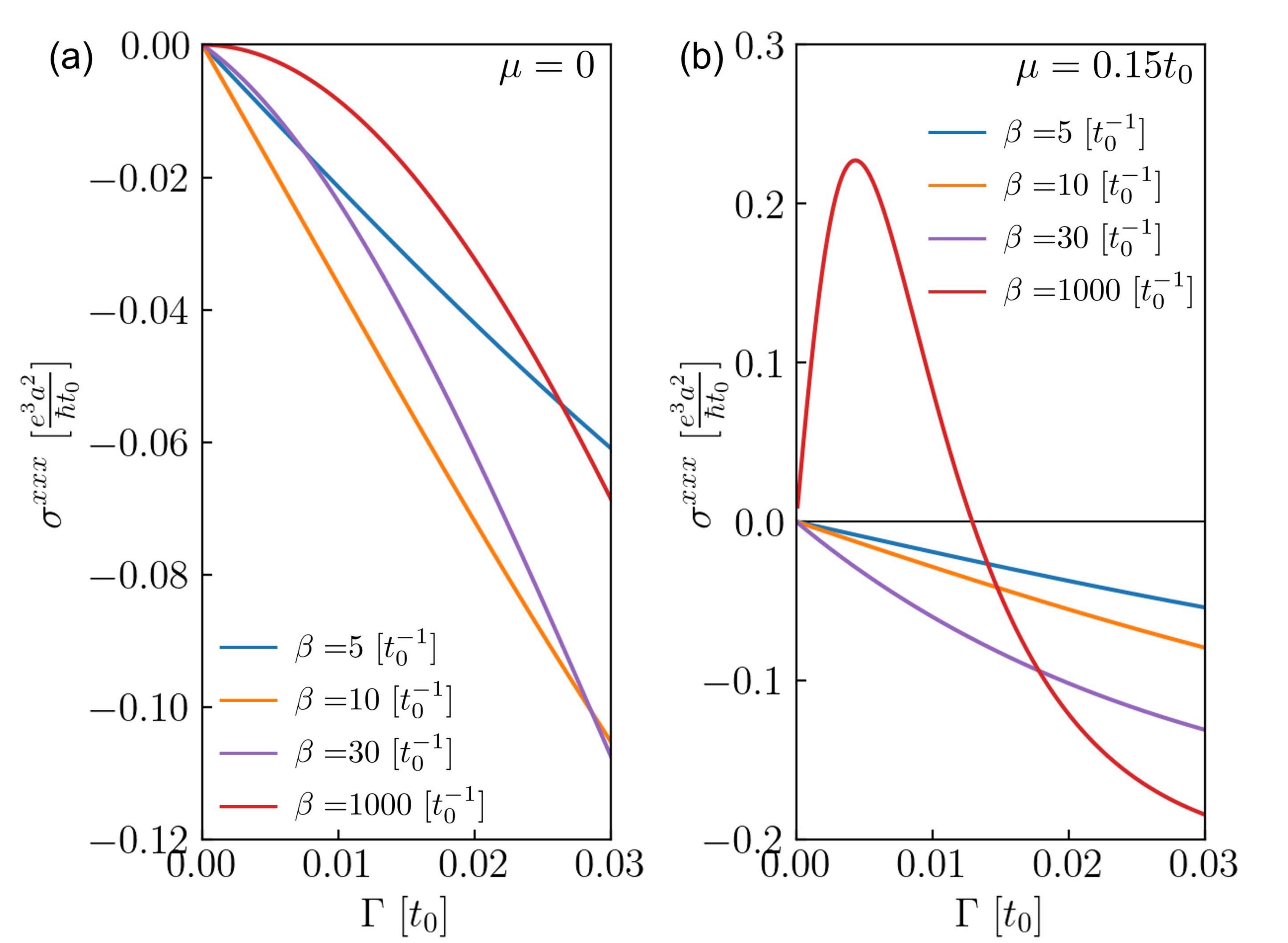}
  \caption{Nonreciprocal current $\sigma^{xxx}$ of the Rice--Mele model as a function of relaxation rate $\Gamma$. The nonreciprocal current is plotted for different values of the inverse temperature $\beta$. (a) Nonreciprocal current in the semiconducting case with $\mu=0$. (b) Nonreciprocal current in the metallic case with $\mu=0.15t_0$.}
  \label{fig:riceMelevsGamma}
\end{figure}

Figure~\ref{fig:riceMelevsGamma} (a) shows the nonreciprocal current $\sigma^{xxx}$ as a function of relaxation rate $\Gamma$ for the semiconducting case ($\mu=0$).
In the low temperature regime $\beta \Gamma \gg 2\pi$, the nonreciprocal current shows a quadratic dependence on $\Gamma$ for small $\Gamma$, as described by Eq.~\eqref{eq:lowTmpGammaSquared}.
In the high temperature regime $\beta \Gamma \ll 2\pi$, the nonreciprocal current shows a linear dependence on $\Gamma$ for small $\Gamma$, as described by Eq.~\eqref{eq:highTmpGammaLinear}.
On the other hand, Fig.~\ref{fig:riceMelevsGamma} (b) shows the nonreciprocal current $\sigma^{xxx}$ as a function of relaxation rate $\Gamma$ with the chemical potential in the conduction band $\mu=0.15t_0$.
In this case, the nonreciprocal current shows a linear dependence on $\Gamma$ for small $\Gamma$ at any temperature.

\section{Numerical calculation for three band model}\label{app:threeBand}
\begin{figure}
  \includegraphics[width=\linewidth]{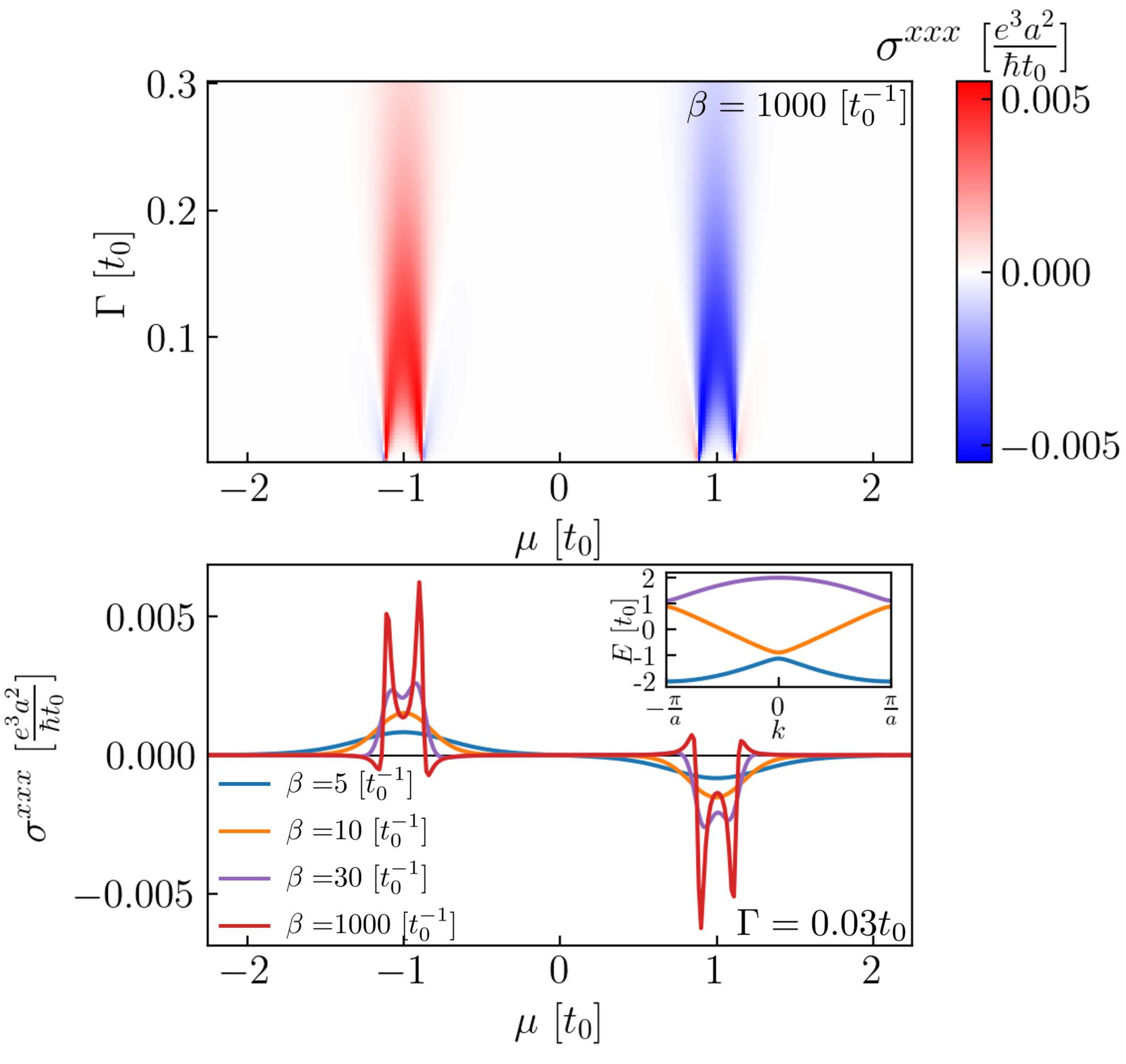}
  \caption{Nonreciprocal current $\sigma^{xxx}$ of the three band model. (a) Color plot of $\sigma^{xxx}$ as a function of chemical potential $\mu$ and relaxation rate $\Gamma$. (b) $\sigma^{xxx}$ as a function of $\mu$ for different values of $\beta$. The inset of (b) shows the band structure of the three band model.}
  \label{fig:threeBand}
\end{figure}

In this section, we show the result of the numerical calculation of Eq.~\eqref{eq:sigma2} for a three band model that is given as, 
\begin{align}
  H=\sum_{i}t_1 c_{Ai}^\dagger c_{Bi}+t_2 c_{Bi}^\dagger c_{Ci}+t_3 c_{Ci}^\dagger c_{Ai+1}+h.c.,
\end{align}
where $c_{Ai}$, $c_{Bi}$, and $c_{Ci}$ are the annihilation operators of the $A$, $B$, and $C$ sublattices at the $i$-th unit cell, respectively.
The hopping parameters are set to $t_1=1.1t_0$, $t_2=t_0$, and $t_3=0.9t_0$.
The bond lengths between the $A$ and $B$ sublattices, the $B$ and $C$ sublattices, and the $C$ and $A$ sublattices are set to $a/3$, where $a$ is the lattice constant.  
This model has time-reversal symmetry and lacks inversion symmetry, which allows for a finite nonreciprocal current arising from the present mechanism.
Figure~\ref{fig:threeBand} (a) shows the color plot of the nonreciprocal current $\sigma^{xxx}$ as a function of chemical potential $\mu$ and relaxation rate $\Gamma$.
The nonreciprocal current is enhanced when the chemical potential lies near the band gap and the relaxation rate is comparable to the band gap.
Figure~\ref{fig:threeBand} (b) shows the nonreciprocal current $\sigma^{xxx}$ as a function of chemical potential $\mu$ for different values of inverse temperature $\beta$.
In the low temperature regime $\beta \Gamma \gg 2\pi$, the nonreciprocal current shows peaks and sign changes near the band edge, which is attributed to the $f_a''$ weight in Eq.~\eqref{eq:skewness} or the $\frac{d n_a}{d\mu}$ weight in Eq.~\eqref{eq:skewness-low-temp}.
In the high temperature regime $\beta \Gamma \ll 2\pi$, the nonreciprocal current shows a single peak in the band gap because of the difference between the Fermi distribution functions of the conduction and valence bands.
These features are the same as those of the Rice--Mele model discussed in the main text.

\section{Order estimation} \label{app:orderEstimation}
In this section, we estimate the order of nonreciprocity induced by the present mechanism.
When the frequency of the applied electric field $\omega$ is sufficiently small, the electric current density induced by the applied electric field $E_x(t)=E_x e^{-i\omega t}+c.c.$ is given by 
\begin{align}
  j_x(t)=&\sigma^{xx} E_x(t)  +\sigma^{xxx} E_x^2(t) +\cdots.
\end{align}
The electric current $I$ is given by
\begin{align}
  I_x\equiv& Sj_x(t)=\sigma^{xx}2\frac{V_x S}{l}\cos\omega t+\sigma^{xxx}(2+2\cos 2\omega t)\frac{V_x^2}{l^2}S+\cdots
\end{align}
where $S$ is the cross-sectional area of the sample, $l$ is the length of the sample, and $V_x=E_x l$ is the voltage applied to the sample.
By defining linear resistance $R^\omega \equiv l/(\sigma^{xx} S)$, we can rewrite the above equation as
\begin{align}
  IR^\omega =& 2V_x \cos\omega t +2\frac{\sigma^{xxx}}{\sigma^{xx}}\frac{V_x^2}{l} (1+\cos 2\omega t)+\cdots \nonumber \\
  \equiv & 2V_x \cos\omega t +V^0 +V^{2\omega} \cos 2\omega t+\cdots.
\end{align}
Here, $V^{2\omega} \equiv 2\frac{\sigma^{xxx}}{\sigma^{xx}}\frac{V_x^2}{l}$ is the second harmonic voltage.
Using the amplitude of the first-harmonic current $\sigma^{xx}2\frac{V_x S}{l}$, we can define the second harmonic resistance $R^{2\omega}$ as $R^{2\omega} \equiv V^{2\omega}/(\sigma^{xx}2\frac{V_x S}{l})=\frac{\sigma^{xxx}}{(\sigma^{xx})^2}\frac{V_x}{S}$.
$V^{2\omega}$ and $R^{2\omega}$ are often used as measures of nonreciprocity in experiments~\cite{Ideue2017,Yokouchi2017,Wakatsuki2017,Yasuda2019,Itahashi2020,Zhang2020,Nakamura2025,Li2021,Wang2022,Wakamura2024}.

Since $V^{2\omega}$ and $R^{2\omega}$ depend on $\sigma^{xx}$, we derive linear conductivity $\sigma^{xx}$ taking into account the same relaxation process as the nonlinear conductivity.
The linear conductivity $\mathcal{K}^{\mu\alpha}(\omega)$ is defined by $j_\mu^{(1)}(t)=\sum_\alpha \int \frac{d\omega}{2\pi } \mathcal{K}^{\mu\alpha} (\omega) A_\alpha (\omega) e^{-i\omega t}$.
Using Matsubara formalism, $\mathcal{K}^{\mu\alpha}(\omega)$ is given as
\begin{align}
    \mathcal{K}^{\mu\alpha}(\omega)=&-\left(\frac{e}{\hbar}\right)^2\frac{-1}{2\pi i}\int dx f(x) \mathrm{tr}\left[(G^R(x)-G^A(x))H^{\mu\alpha}\right. \nonumber \\
    &\hspace{1.5cm}+H^\mu G^R(x+\hbar\omega)H^\alpha (G^R(x)-G^A(x)) \nonumber \\
    &\hspace{1.5cm}\left. +H^\mu (G^R(x)-G^A(x)) H^\alpha G^A(x-\hbar\omega) \right]\nonumber \\
    =&\hbar\omega\left(\frac{e}{\hbar}\right)^2\frac{2i\Gamma}{2\pi i}\int dx f \mathrm{tr}[H^\mu G^{R}G^R(x+\hbar\omega)H^\alpha  G^R G^A \nonumber \\
    &\hspace{2cm}- H^\mu  G^R G^A H^\alpha G^{A}G^A(x-\hbar\omega)] .
\end{align}
Here, we used $-2i\Gamma G^R(x) G^A(x) =G^R(x)-G^A(x)$ and $\int \frac{d^d k}{(2\pi)^d}\partial_\alpha \mathrm{tr}[G^R G^A H^{\mu}]=0$.
Similar to the nonlinear conductivity, $\mathcal{K}^{\mu\alpha}(\omega)=O(\omega)$ and we can define the linear conductivity as $\sigma^{\mu\alpha}=-i\partial_\omega \mathcal{K}^{\mu\alpha}(\omega)_{\omega=0}$, which is given as
\begin{align}
  \sigma^{\mu\alpha}(\omega)
  =& \frac{e^2}{\hbar}\frac{2\Gamma}{2\pi i}\int dx \nonumber \\
  &\quad \times \mathrm{tr}[(f(x)\partial_\mu G^R -f(x+\hbar\omega)\partial_\mu G^A(x+\hbar\omega))\nonumber \\
  &\hspace{3cm} \times G^R(x+\hbar\omega)H^\alpha G^A(x)].
\end{align}
Therefore, the static linear conductivity is given as
\begin{align}
  \sigma^{\mu\alpha}
  =& \frac{e^2}{\hbar}\frac{2\Gamma}{2\pi i}\int dx f(x)\mathrm{tr}[G^R H^\mu G^R G^R H^\alpha G^A]+c.c.\nonumber \\
  =& \frac{e^2}{\hbar} 2\Gamma \int \frac{d^d k}{(2\pi)^d} \sum_{ab} H_{ab}^\mu H_{ba}^\alpha I^{(4)}_{ab} +c.c., 
\end{align}
with
\begin{align}
  I^{(4)}_{ab}=&\frac{1}{2\pi i}\int dx f(x) \frac{1}{(x-\xi_a+i\Gamma)(x-\xi_b+i\Gamma)^2(x-\xi_a-i\Gamma)} \nonumber \\
  I^{(4)}_{a\neq b}=&f_+(\xi_a+i\Gamma)\frac{1}{2i\Gamma(\xi_{ab}+2i\Gamma)^2} - f_-'(\xi_b-i\Gamma)\frac{1}{\xi_{ba}(\xi_{ba}-2i\Gamma)} \nonumber \\
  &+f_-(\xi_b-i\Gamma)\left[
    \frac{1}{\xi_{ba}^2(\xi_{ba}-2i\Gamma)}+\frac{1}{\xi_{ba}(\xi_{ba}-2i\Gamma)^2}
  \right]\nonumber \\
  &\qquad -f_-(\xi_a-i\Gamma)\frac{1}{\xi_{ab}^2(-2i\Gamma)}\nonumber \\
  I^{(4)}_{aa}=&\frac{f_+(\xi_a+i\Gamma)}{(2i\Gamma)^3} -\frac{1}{2}\frac{f_-''(\xi_a-i\Gamma)}{(-2i\Gamma)}+\frac{f_-'(\xi_a-i\Gamma)}{(-2i\Gamma)^2}-\frac{f_-(\xi_a-i\Gamma)}{(-2i\Gamma)^3}
\end{align}
For longitudinal conductivity, we set $\mu=\alpha$ and obtain,
\begin{align}
  \sigma^{\alpha\alpha}
  =& \frac{e^2}{\hbar} \int \frac{d^d k}{(2\pi)^d} \left\{\sum_{a\neq b} H_{ab}^\alpha H_{ba}^\alpha \left[-\frac{2\Gamma f_+'(\xi_a+i\Gamma)}{\xi_{ab}(\xi_{ab}+2i\Gamma)}+c.c.\right]\right. \nonumber \\
  &\left. +\sum_a H_{aa}^\alpha H_{aa}^\alpha \left[-\frac{f_+''(\xi_a+i\Gamma)}{2i}-\frac{f_+'(\xi_a+i\Gamma)}{2\Gamma}+c.c.\right]\right\}.
\end{align}

\begin{figure}[t]
  \centering
\includegraphics[width=\linewidth]{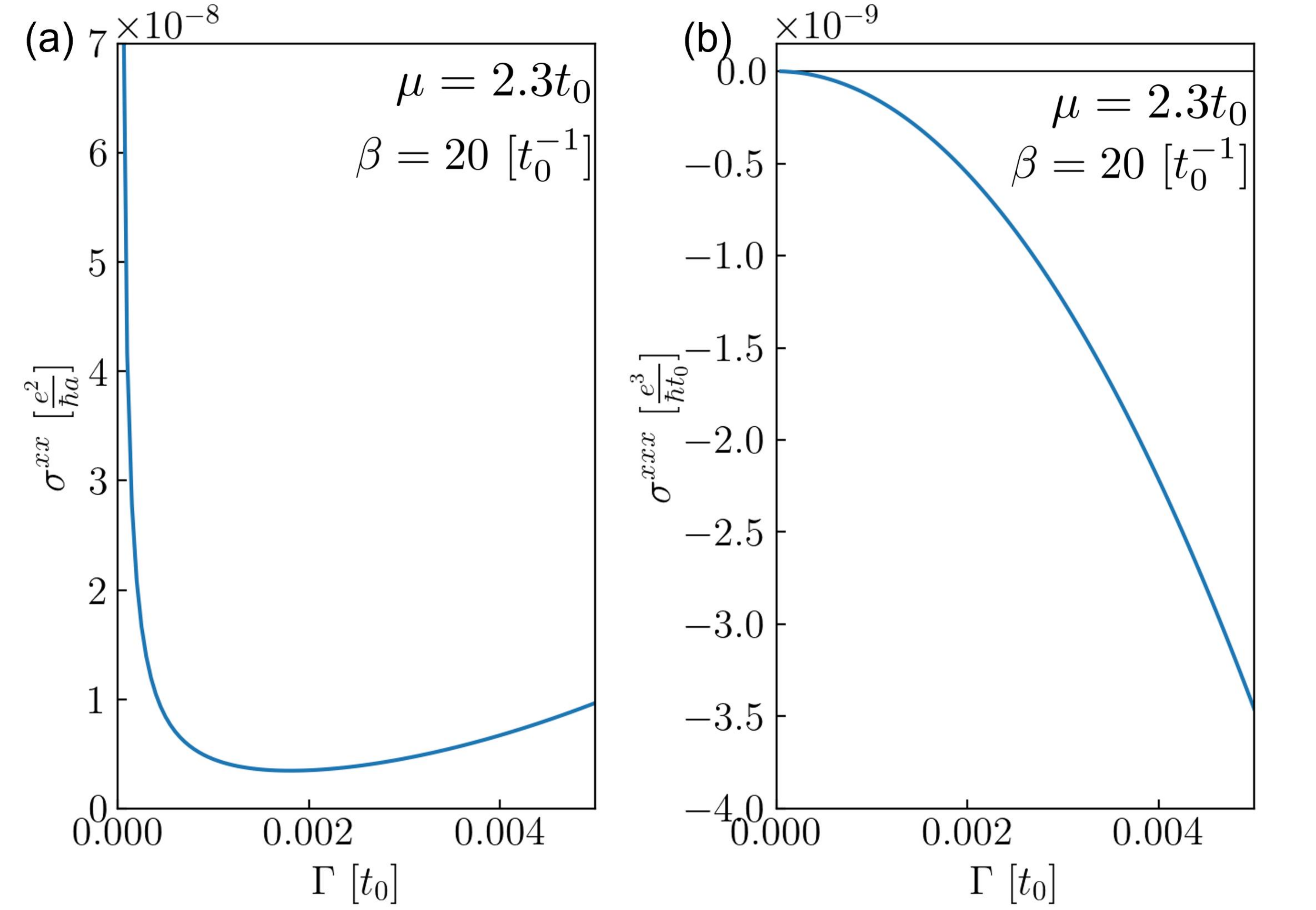}
  \caption{(a) The linear conductivity and (b) the nonlinear conductivity as a function of the relaxation rate $\Gamma$. We set the parameters as $m=3.1t_0,\delta t=0.3t_0$, and $t_1=5.0t_0$.}
  \label{fig:orderVsGamma}
\end{figure}

Figure~\ref{fig:orderVsGamma} shows the linear and nonlinear conductivities as a function of the relaxation rate $\Gamma$ for a three-dimensional extension of the Rice--Mele model $\mathcal{H}(k)=t_0 \cos k_x a\sigma_x +\delta t \sin k_xa \sigma_y +(m+t_1 (2-\cos k_y a-\cos k_z a))\sigma_z$ with $m=3.1t_0$, $\delta t=0.3t_0$ and $t_1=5.0t_0$.
We set the chemical potential $\mu=2.3t_0$ and the temperature $\beta=20/t_0$.
For the representative value $t_0=0.5\,\mathrm{eV}$ used below, this corresponds to $T=290\,\mathrm{K}$, which is in the room temperature range.
These parameters are chosen such that the system is a nonmagnetic doped semiconductor with broken inversion symmetry.
Although the large band gap suppresses the nonlinear conductivity $\sigma^{xxx}$, it also suppresses the linear conductivity $\sigma^{xx}$ and thus reduces Joule heating.
This situation is suitable for observing the second-harmonic voltage $V^{2\omega}$ and the second-harmonic resistance $R^{2\omega}$ induced by the present mechanism.

To estimate the signal magnitude in physical units, we adopt representative parameter values $t_0=0.5\,\mathrm{eV}$, $a=0.3\,\mathrm{nm}$, $S=1\,\mathrm{mm}^2$, and $l=1\,\mathrm{mm}$, where $S$ and $l$ denote the cross-sectional area and the sample length, respectively.
With these parameters, the electron occupation density of the conduction band is estimated from the spectral function as $n=\frac{1}{a^3}\int \frac{dk^3}{(2\pi)^3}\int dx f(x)(-\frac{1}{\pi}\mathrm{Im}\frac{1}{x+\mu-\epsilon_{\mathrm{up}}+i\Gamma})=3.2\times 10^{18}\, [\mathrm{cm}^{-3}]$, where $\epsilon_{\mathrm{up}}$ is the eigenenergy of the upper band.
The hole density in the valence band is also of the same order of magnitude.
We consider the case of a moderate dissipated power of $1\,\mathrm{mW}$, which leads to the applied voltage $V_x$ of $\simeq 10\,\mathrm{V}$ and the first-harmonic current amplitude of $\simeq0.05\,\mathrm{mA}$.
In this setup along with $\Gamma=0.002t_0$, (corresponding to the relaxation time $\tau \approx 0.3\,\mathrm{ps}$,) the second-harmonic voltage $V^{2\omega}$ is approximately $-0.02\,\mathrm{mV}$ and the second-harmonic resistance $R^{2\omega}$ is approximately $-0.3\,\Omega$.
These values of $V^{2\omega}$ and $R^{2\omega}$ are feasible for experimental observation in nonmagnetic polar semiconductors with low carrier density~\cite{Ideue2017,Yokouchi2017,Wakatsuki2017,Yasuda2019,Itahashi2020,Zhang2020,Nakamura2025,Li2021,Wang2022,Wakamura2024}.

\bibliographystyle{apsrev4-1}
\bibliography{references}

\end{document}